\documentclass[review]{elsarticle}

\usepackage{natbib}
\usepackage{lineno, hyperref}

\usepackage{wrapfig}

\usepackage{amssymb,amsmath}
\usepackage{bm,upgreek}
\usepackage{mathrsfs}
\usepackage{graphics}
\usepackage{amsfonts}
\usepackage{color}

\usepackage[symbol]{footmisc}

\usepackage{lipsum}

\graphicspath{{Fig3D/}{Fig/}}

\newcommand\rd{{\rm{d}}}
\newcommand{\bfm}[1]{{\rm\bf #1}}

\definecolor{myviolet}{RGB}{255,0,255}
\definecolor{darkgreen}{rgb}{0,0.5,0}
\definecolor{shade}{RGB}{176,176,176}

\journal{International Journal of Mechanical Sciences}

\begin{document}

\begin{frontmatter}

\title{
Indentation-induced martensitic transformation in SMAs: insights from phase-field simulations$^\dagger$
}

\author[IPPT]{Mohsen Rezaee-Hajidehi}
\ead{mrezaee@ippt.pan.pl}

\author[FMP]{Karel T\r{u}ma}
\ead{ktuma@karlin.mff.cuni.cz}

\author[IPPT]{Stanis{\l}aw Stupkiewicz\corref{cor1}}
\ead{sstupkie@ippt.pan.pl}

\cortext[cor1]{Corresponding author.}

\address[IPPT]{Institute of Fundamental Technological Research (IPPT), Polish Academy of Sciences,\\
Pawi\'nskiego 5B, 02-106 Warsaw, Poland.}

\address[FMP]{Faculty of Mathematics and Physics, Charles University, Sokolovsk\'a 83, 186 75, Prague, Czech Republic}

\begin{abstract}
Direct experimental characterization of indentation-induced martensitic microstructures in pseudoelastic shape memory alloys (SMAs) is not possible, and thus there is a lack of evidence and understanding regarding the microstructure pattern and related features. To fill this gap, in this work we employ the phase-field method to provide a detailed and systematic analysis of martensitic phase transformation during nanoindentation. A recently-developed finite-element-based computational model is used for this purpose, and a campaign of large-scale 3D simulations is carried out. 
First, the orientation-dependent indentation response in CuAlNi (a widely studied SMA) is examined. A detailed investigation of the predicted microstructures reveals several interesting features, some of them are consistent with theoretical predictions and some can be (to some extent) justified by experiments other than micro/nanoindentation. The results also highlight the key role of finite-deformation effects and elastic anisotropy of the phases on the model predictions. Next, a detailed study of indentation-induced martensitic transformation in NiTiPd (a potential low-hysteresis SMA) with varying Pd content is carried out. In terms of hysteresis, the results demonstrate the prevailing effect of the transformation volume change over phase compatibility in the conditions imposed by nanoindentation and emphasize on the dominant role of the interfacial energy at small scales. Results of such scope have not been reported so far.\footnotetext[2]{Published in \emph{Int. J. Mech. Sci.}, \textbf{245}, 108100, 2023, doi: 10.1016/j.ijmecsci.2023.108100}
\end{abstract}


\end{frontmatter}

\section{Introduction}
Shape memory alloys (SMAs) are a class of smart materials that have gained tremendous attention in research and industry mainly owing to their unique capabilities of pseudoelasticity and shape memory effect \citep{OtsukaWayman98,mohdjani20141078}, which stem from the crystallographically reversible martensitic transformation \citep{Bhattacharya2003}. With the increasing development of SMA miniaturized applications \citep{kohl2004shape,bellouard2008582,miyazaki2009thin,chluba2015ultralow}, it is of utmost importance to deepen the understanding of the mechanical behaviour of SMAs at micro/nano-scale regimes \citep{karami2021100141}. One of the popular experimental techniques in this context is the instrumented micro/nanoindentation  \citep{schuh2006nanoindentation, FischerCripps2011}. Over the last two decades, the micro/nanoindentation technique has been extensively applied to SMA materials to characterize, for instance, the small-scale mechanical properties \citep{frick2006stress,zhang2006nanoscale,kan2013oliver, laplanche2014orientation}, size effects \citep{amini2011depth,montecinos2021indentation}, and microstructural changes induced by martensitic transformation \citep{zheng2008tem,pfetzing2014direct}.

In general, the interpretation of the indentation tests, apart from the measured load--indentation depth curve, relies upon the surface profile of the residual imprint \citep{bolzon20042957} and the post-mortem microstructural analysis \citep{rester2007microstructural}. In SMAs, using advanced imaging techniques, such as atomic force microscopy (AFM), e.g., \citep{shaw2003shape}, scanning electron microscopy (SEM), e.g., \citep{laplanche2014sudden}, or transmission electron microscopy (TEM), e.g., \citep{zheng2008tem,pfetzing2014direct}, valuable information regarding the deformation behaviour and the microstructure can be obtained. Chief among them, TEM in combination with electron diffraction measurements have been utilized to analyze the characteristic changes in the remnant microstructure beneath the indenter and to identify the mechanism underlying the inelastic deformation. However, TEM method is only operative on thin foils, the microstructure of which may not genuinely represent the martensitic microstructure in bulk materials \citep{laplanche2014sudden}. At the same time, AFM and SEM micrographs can only reflect the surface morphology of the sample and deliver no information about the through-depth microstructure. 

On top of the limitations and challenges imposed by the measurement techniques, one major issue in the indentation testing of SMAs arises when the material is in the pseudoelastic state. Different from conventional elastoplastic metals or self-accommodated (pseudoplastic) SMAs, the inelastic deformation and the evolved martensitic microstructure disappear (completely or partially) during unloading as a result of the reverse transformation, especially in the case when shallow spherical indents are made \citep{ni2003microscopic,li2015724}. Accordingly, since no remnant microstructure exists, material characterization is mostly based on the analysis of the load--indentation depth response. In this respect, attempts have been made to probe the indentation-induced microstructure, e.g., via tuning the alloy composition such that both austenite and martensite phases are stable at room temperature, and thereby reverse transformation is suppressed during unloading \citep{pfetzing2014direct,laplanche2014orientation}. It is also noteworthy to point out that the stress-induced phase transformation in nanoindentation experiments is typically influenced by the inevitable generation of dislocation plasticity \citep{wood2006measurement, pfetzing2010nanoindentation,kumar2020assessment}, which may occur in view of the large strains and strain gradients beneath the indenter (typically, in the case of sharp Berkovich tip, or spherical tips at high loads). Thus, restricting the dislocation plasticity represents another major experimental issue in the study of martensitic phase transformation in SMAs subject to nanoindentation. We conclude this discussion by stressing that no experimental evidence of indentation-induced martensitic microstructure, in the sense of 3D spatial arrangement of martensite variants, has been reported in the literature to date, in particular, for pseudoelastic SMAs.

In light of the experimental issues discussed above, modeling can be viewed as a cornerstone to gain a deeper understanding of the martensitic microstructure in nanoindentation. Existing modeling approaches in this area cover a broad range of spatial scales, spanning from atomistic up to macroscopic scale. Atomistic models, e.g., those based on the molecular dynamics (MD) techniques, while able to rigorously resolve the atomic-level interactions within the material phases, are limited to very small spatial scales (up to few tens of nanometers) and temporal scales (up to hundreds of picoseconds), e.g., \citep{pfetzing2013crystallographic,chen2018molecular,doan2020influences,ko2021atomistic}. Micromechanical crystal-plasticity-like models \citep{pan2007multi, dhala2019analyses,hossain2021finite} and macroscopic models \citep{wood2006measurement,yan2007analysis, amini2013phase}, on the other hand, are applicable at higher scales, i.e., scales at which the spatial heterogeneity of the martensite phase (including the formation of different martensite variants and the associated phase boundaries) is justifiably not taken into consideration. 

Meso-scale modeling using the phase-field method can be regarded as the most suitable modeling strategy for the problem at hand. The phase-field method is a powerful and ubiquitous modeling tool that is based on the notion of diffuse interfaces and enables a detailed description of the microstructural features. The method has thus been widely used to study the microstructure evolution in various material systems, see \citep{TOURRET2022100810} for a recent review. Among them, evolution of martensitic microstructure has been the subject of a substantial number of phase-field models, including those implemented using the spectral (FFT-based) methods, e.g., \citep{wang1997three,artemev2000three, jin2001three,ahluwalia2004landau,shu2008multivariant, zhong2014phase,zhao2020finite}, and the finite-element method, e.g., \citep{levitas2009displacive,hildebrand2012phase,yeddu2012three, tuuma2016size,cisse2021transformation,tuuma2021phase,xu2020phase}.

There seem to be, however, only a few applications of the phase-field method to nanoindentation problems \citep{clayton2011msmse, basak2019finite,qi2022phase}. This short list includes the authors' previous work on multivariant martensitic transformation \citep{rezaee2020phase} and its extension to account for rate-independent dissipation effects \citep{rezaee2021micromorphic}. Note that all the five studies referenced above are restricted to 2D analyses, mainly due to the high computational cost and complexity associated with 3D computations.

While there has been a significant progress in the development of phase-field models suitable for large-scale 3D computations, the majority of the models are tailored to the popular FFT-based spectral solvers. Such models, despite being computationally highly efficient, are in general limited to a periodic unit cell, and thus complex geometries and arbitrary boundary conditions (including contact interactions) cannot be applied. On the other hand, while the finite-element method is not restricted by the geometry and boundary conditions, finite-element-based models able to cope with large-scale 3D computations have been scarcely reported. In the present context, as far as we could ascertain, the only existing models are those in \citep{yeddu2012three} and \citep{dhote201548}, the latter employing the framework of isogeometric analysis.

In our recent work \citep{tuuma2021phase}, we have developed a phase-field-based computational model for 3D analysis of martensitic microstructure in SMAs. The model is characterized by the following features: (i) the constitutive description is formulated within the finite-strain framework with the Hencky-type anisotropic elastic strain energy and the double-obstacle potential defining the interfacial energy, (ii) the variational formulation is based on the incremental energy minimization framework, (iii) the penalty method is used to regularize the inequality constraints imposed on the order parameters (note the double-obstacle potential), as well as those related to unilateral contact (note the contact interaction with the indenter), and finally (iv) the finite-element method is used for the spatial discretization, and this is combined with an efficient parallelization (so that large-scale problems can be effectively solved). While individually none of the above features is truly distinctive as compared to other existing phase-field models, it is the combination of all these features that makes our computational model suitable for 3D analysis of microstructure evolution in nanoindentation. The model can thus be considered unique, as it represents the only application of the phase-field method to this class of problems.

The focus of our previous work \citep{tuuma2021phase} has been mostly on the computational aspects. In particular, basic features and capabilities of the model have been demonstrated through the analysis of the microstructure evolution during nanoindentation and its computational robustness has been verified. At the same time, no detailed analysis of the microstructure evolution and mechanical behaviour has been attempted.
 
The same model is employed in this work with the main aim of addressing two specific problems. The first problem concerns the effect of crystal orientation on the nanoindentation response in a CuAlNi single crystal (a widely studied SMA material with well-documented material parameters). This effect has been investigated experimentally (though for a different SMA material, namely NiTi) and a strong orientation dependence of the mechanical response and the residual imprints has been found \citep{pfetzing2013crystallographic,laplanche2014orientation,laplanche2014sudden}. The findings, however, could not be supported by the evidence of martensitic microstructure evolution, and only recently some indications have been given by MD simulations \citep{ko2021atomistic}. The second problem is motivated by the peculiar martensitic transformation behaviour in low-hysteresis SMAs. Certain compositions in these alloys give rise to special (twinless) microstructural patterns and lead to a very low transformation hysteresis \citep{cui2006combinatorial,delville2010transmission, evirgen2016relationship}, a feature that is of critical importance for numerous practical applications \citep{gu2018phase}. Upon exploiting the nanoindentation setting, the goal is to study this aspect in NiTiPd single crystal with varying Pd content. As far as we are aware, the only reported modeling study on this matter is that by \citet{salman2012role} who studied microstructure evolution driven by the self-accommodation mechanism (in the absence of external loads) in NiTiPd by using a small-strain phase-field model. Finally, we also show the impact of some characteristic model features on the simulation results and highlight their key roles in effectively predicting the microstructure evolution.

\section{Phase-field model for nanoindentation problem}\label{sec-PFmodel}
In this section, we present the computational model that will be next used for the simulation of indentation-induced martensitic transformation in SMA single crystals. The finite-strain phase-field model of multivariant martensitic transformation is introduced in Section~\ref{sec-modelformul} and its finite-element implementation and computational treatments are briefly discussed in Section~\ref{sec-FEimp}. A more detailed description of the model can be found in \citet{tuuma2021phase}. Finally, in Section~\ref{sec-setup}, the setup of the indentation problem and the material parameters adopted in the simulations are discussed.
\subsection{Model formulation}\label{sec-modelformul}
Within the phase-field method, each phase $i$ is characterized by a continuous variable called order parameter, denoted here by $\eta_i$. Considering austenite (parent phase) and $N$ variants of martensite, the model must account for $N+1$ order parameters, which are subject to the following constraints
\begin{equation}\label{Eq-constrains}
0 \leq \eta_i \;\; \text{for} \; i=0,\dots,N \quad \text{and} \quad \sum_{i=0}^N \eta_i=1.
\end{equation}
In the present setting, the order parameters can be understood as the phase volume fractions, and are used to interpolate the material properties across the diffuse interfaces. An alternative approach is to use the order parameters to construct special interpolating polynomials \citep{clayton2011msmse, levitas2018phase}, typically, jointly with the so-called double-well potential, rather than the double-obstacle potential \citep{steinbach2009phase} used here for the interfacial energy, cf.~Eq.~\eqref{Eq-int}.

The deformation gradient $\bfm{F}=\nabla \bm{\upvarphi}$, with $\bm{\upvarphi}$ denoting the deformation mapping between the reference and the current configurations, is the primary kinematic variable in the finite-strain theory, and admits a multiplicative decomposition into the elastic part $\bfm{F}^\text{e}$ and the transformation part $\bfm{F}^\text{t}$, i.e., $\bfm{F}=\bfm{F}^\text{e}\bfm{F}^\text{t}$. In general, there is some flexibility in the definition of the transformation part $\bfm{F}^\text{t}$, see \citep{tuuma2016size,basak2017interfacial} for the related discussion. In the present model, $\bfm{F}^\text{t}$ is defined as a linear combination of the transformation stretch (Bain strain) tensors $\bfm{U}^\text{t}_i$, i.e.,
\begin{equation}\label{Eq-Ft}
\bfm{F}^\text{t}=\sum_{i=0}^N \eta_i \bfm{U}_i^\text{t}.
\end{equation}
As usual, the undeformed austenite is taken as the reference configuration, thus for the pure austenite state, i.e., $\eta_0=1$, we have $\bfm{F}^\text{t}=\bfm{U}^\text{t}_0=\bfm{I}$, where $\bfm{I}$ is the second-order identity tensor. The transformation stretches $\bfm{U}_i^\text{t}$ ($i=1,\dots,N$) are known from the crystallography.

Isothermal conditions are assumed throughout this work. While thermomechanical coupling effects are known to be important in SMAs (at least at the macroscopic scale) \citep{shaw1995thermomechanical,zhang2010experimental,yin2021thermomechanical}, they are often neglected in phase-field modeling of martensitic transformation \citep{shu2008multivariant,levitas2009displacive,yeddu2012three,hildebrand2012phase}. In general, incorporation of thermomechanical couplings in the phase-field framework does not pose any major difficulty, however, it is out of the scope of the current study.

The total Helmholtz free energy $\mathcal{F}$ is defined as the sum of the chemical energy $F_\text{chem}$, elastic strain energy $F_\text{el}$ and interfacial energy $F_\text{int}$, integrated over the entire body domain $B$,
\begin{equation}
\mathcal{F}=\int_B \left(F_\text{chem}+F_\text{el}+F_\text{int} \right) \rd V.
\end{equation}
The chemical energy is expressed as
\begin{equation}\label{Eq-chem}
F_\text{chem}=\sum_{i=0}^N \eta_i F_i^0,
\end{equation}
with $F_i^0$ as the chemical energy related to the phase $i$ (constant in isothermal conditions). The elastic strain energy $F_\text{el}$ is adopted as a quadratic function of the elastic logarithmic strain $\bfm{H}^\text{e}$,
\begin{equation}\label{Eq-elasticEn}
F_\text{el}
=\frac{1}{2}(\det \bfm{F}^\text{t})\bfm{H}^\text{e} \cdot \mathbb{L}\bfm{H}^\text{e}, \quad \bfm{H}^\text{e}=\frac{1}{2}\log \bfm{C}^\text{e},
\end{equation}
where $\bfm{C}^\text{e}=(\bfm{F}^\text{e})^\text{T}\bfm{F}^\text{e}$ is the elastic right Cauchy-Green tensor, and $\mathbb{L}=\sum_{i=0}^N \eta_i \mathbb{L}_i$ is the fourth-order elastic stiffness tensor obtained by Voigt-type averaging of the stiffness tensors $\mathbb{L}_i$ of individual phases. The choice of the logarithmic strain energy \eqref{Eq-elasticEn} is motivated by its superior performance compared to the popular St.~Venant--Kirchhoff model (formulated in terms of the elastic Green strain tensor $\bfm{E}^\text{e}=\frac{1}{2}(\bfm{C}^\text{e}-\bfm{I})$), in particular, under large compressive stresses, which is the case in the indentation problem studied in this paper. This issue is discussed in detail in \citep{rezaee2021note}.

Finally, the double-obstacle potential \citep{steinbach2009phase} is adopted for the interfacial energy,
\begin{equation}\label{Eq-int}
F_\text{int}=\sum_{i=0}^N \sum_{j=i+1}^N \frac{4 \gamma_{ij}}{\pi \ell_{ij}} \left( \eta_i \eta_j - \ell_{ij}^2 \nabla \eta_i \cdot \nabla \eta_j \right),
\end{equation}
where $\gamma_{ij}$ and $\ell_{ij}$ denote, respectively, the interfacial energy density (per unit area) and the interface thickness parameter associated with the diffuse interface between phases $i$ and $j$. The interfacial energy in the form \eqref{Eq-int} results in the theoretical (unstressed) interface thickness $\lambda_{ij}=\pi \ell_{ij}$. 
Note that, compared to the more popular double-well potential \citep{steinbach2009phase}, the double-obstacle potential brings several advantages. It leads to an exact elastic response and less diffuse interfaces, and on top of that, it naturally provides a barrier against the occurrence of a spurious third phase within a diffuse binary interface, see the detailed discussion in \citep{tuuma2016phase}. Moreover, when the double-obstacle potential is combined with linear weighting, as in Eqs.~\eqref{Eq-Ft} and \eqref{Eq-chem}, phase nucleation proceeds spontaneously and no special treatment is needed to trigger nucleation.

The formulation of the model is completed with the specification of the global dissipation potential $\mathcal{D}$. A viscous-type dissipation is employed, which is formulated in terms of the rate of the order parameters $\eta_i$, i.e., 
\begin{equation}\label{Eq-diss}
\mathcal{D}=\int_B \left( \sum_{i=0}^N \frac{\dot{\eta}_i^2}{2m_i} \right) \rd V,
\end{equation}
where $m_i$ is the mobility parameter related to the phase $i$. Eq.~\eqref{Eq-diss} implies that the effective mobility of the interface between phases $i$ and $j$ is equal to $m_{ij}=m_i m_j/(m_i +m_j)$ \citep{rezaee2020phase}.

To arrive at the governing equations of the evolution problem, a variational formulation of the model is established following the approach developed by \citet{hildebrand2012phase}, see also \citep{tuuma2016size}. To this end, a global rate-potential $\Pi$ is formulated and is minimized with respect to the rates $\dot{\bm{\upvarphi}}$ and $\dot{\bm{\upeta}}$, where $\bm{\upeta}=\{ \eta_0, \eta_1,\dots, \eta_N \}$. Thus, we have
\begin{equation}\label{Eq-min}
\Pi=\dot{\mathcal{F}}+\mathcal{D} \; \rightarrow \; \min_{\substack{\dot{\bm{\upvarphi}},\dot{\bm{\upeta}}}}
\end{equation}
subject to the constraints specified by Eq.~\eqref{Eq-constrains}. Minimization of the rate-potential $\Pi$ with respect to $\dot{\bm{\upvarphi}}$ implies mechanical equilibrium, while minimization of $\Pi$ with respect to $\dot{\bm{\upeta}}$ yields the evolution equation for the order parameters in the form of the classical Ginzburg--Landau equation, e.g., \citep{penrose1990thermodynamically}. 

In the indentation problem studied in this paper, the load is applied through the contact interaction between the body and the indenter. Hence, the potential of the external loads, which enters the rate-potential $\Pi$ in the general case, vanishes here and is thus absent in Eq.~\eqref{Eq-min}.

\subsection{Finite-element implementation}\label{sec-FEimp}
Details of the computational treatment of the model can be found in \citep{tuuma2021phase}. Below, we briefly describe the most important aspects.

The rate-problem discussed above is transformed to its incremental (finite-step) form by applying the backward Euler scheme. Spatial discretization is performed by the finite-element method, and standard isoparametric four-noded tetrahedral elements with piecewise linear basis functions are used for both the displacements and the order parameters. In the numerical studies reported in Section~\ref{sec-sim}, the CuAlNi and the NiTiPd alloys are considered, both undergoing a cubic-to-orthorhombic phase transformation involving 6 martensite variants. Accordingly, each finite-element node possesses, in addition to 3 displacements, 6 order parameters as global degrees of freedom.

The indentation load is exerted through frictionless contact with a rigid spherical indenter. Unilateral contact conditions are enforced at the nodes of the potential contact surface using the standard penalty method \citep{wriggers2006computational}. Our previous 2D and 3D studies \citep{rezaee2020phase,tuuma2021phase} have shown that the penalty method performs reasonably well. In particular, it has been observed that within a wide range of contact penalty parameters, the computational performance (i.e., the convergence behaviour of the global Newton scheme and of the iterative solver) is not visibly affected. The penalty method is also used to enforce the inequality constraints on the order parameters, see Eq.~\eqref{Eq-constrains}, and it performs similarly well.

The resulting coupled nonlinear equations are solved using the Newton's method and a monolithic solution strategy. Note that, in the context of phase-field modeling, an adequately fine finite-element mesh is required to correctly resolve the diffuse interfaces and capture the subtle features of the evolved microstructure, which naturally leads to large-scale computational problems. As a result, a large system of linear equations has to be solved at each Newton iteration, and this highlights the need for scalable iterative solvers with appropriate preconditioners. Computer implementation is therefore done in Firedrake \citep{rathgeber2016firedrake}, a finite-element package with the access to a variety of solvers and preconditioners available in PETSc library \citep{balay2019petsc}. To efficiently handle the linear problems, we opt to employ the GMRES iterative solver \citep{saad1986gmres} in combination with a geometric multigrid preconditioner \citep{Trottenberg2001}. Multigrid methods exploit a hierarchy of discretizations (in the present implementation, four levels of discretization), and rely on an iterative data projection procedure (prolongation/restriction operators) and smoothing steps across different levels of discretization. For brevity, we skip the detailed description of the multigrid method used, see further details in \citep{tuuma2021phase}. 

All the simulations are performed on the Karolina supercomputer of the IT4Innovations National Supercomputing Center in Ostrava, Czech Republic. Karolina supercomputer consists of 720 standard computing nodes, each equipped with two 64-core AMD EPYC 7H12 processors (2.6 GHz, 256 GB RAM) \citep{Karolina}. Each simulation is run on four computing nodes.

\subsection{Problem setup and material parameters}\label{sec-setup}
In all the simulations, a cuboid computational domain of the size $400\times400\times280$ nm$^3$ is considered. A structured finite-element mesh with approximately 31 million elements (4-noded tetrahedral elements with the edge size of $h=1.9$ nm) and approximately 48 million degrees of freedom is used. The top surface is indented by a spherical indenter of the radius of $R=200$ nm (recall that the indenter is assumed rigid and contact is treated as frictionless). At the same time, the out-of-plane displacements of the lateral and bottom surfaces are restrained. The indentation loading is exerted by prescribing the position of the indenter with the speed of $v=3$ nm/s.

Two SMA materials are considered in the simulations, namely CuAlNi and NiTiPd, both undergoing a cubic-to-orthorhombic transformation
with $N=6$ martensite variants involved. The transformation stretch tensors characterizing the martensite variants 1 and 2 (shown here as a representative case) take the form,
\begin{equation}\label{Eq-stretch}
\bfm{U}_{1,2}^\text{t} =
\begin{pmatrix}
(\alpha+\gamma)/2 & 0 & \pm(\alpha-\gamma)/2 \\
0 & \beta & 0 \\
\pm(\alpha-\gamma)/2 & 0 & (\alpha+\gamma)/2
\end{pmatrix},
\end{equation}
in an orthonormal basis parallel to the axes of the cubic austenite unit cell, with $\alpha$, $\beta$ and $\gamma$ denoting the stretch parameters. The transformation stretch tensors associated with the other variants, $\bfm{U}_{3}^\text{t},\dots,\bfm{U}_{6}^\text{t}$, are obtained by applying the proper rotations from the symmetry point group of the cubic austenite to $\bfm{U}_{1}^\text{t}$, see Tables~3.1 and 4.2 in \citep{Bhattacharya2003}.

Concerning CuAlNi, the stretch parameters are adopted as $\alpha=1.0619$, $\beta=0.9178$ and $\gamma=1.0230$ \citep{otsuka1974morphology}. Full elastic anisotropy of austenite and martensite phases is accounted for, where the elastic constants of the cubic austenite phase ($c_{11}=142$, $c_{44}=96$, $c_{12}=126$, all in GPa) and orthorhombic martensite phase ($c_{11}=189$, $c_{22}=141$, $c_{33}=205$, $c_{44}=54.9$, $c_{55}=19.7$, $c_{66}=62.6$, $c_{12}=124$, $c_{13}=45.5$, $c_{23}=115$, all in GPa) are taken from the available experimental data \citep{suezawa1976behaviour,YasunagaMeasurement}. 

Concerning NiTiPd, several compositions of varying Pd content (Ni$_{50-x}$Ti$_{50}$Pd$_x$, with $x$ denoting the Pd content ranging from 7 to 25 at.\,\%) are considered. The crystallographic lattice parameters of the austenite and martensite phases, and thus the stretch parameters, vary depending on the Pd content. The corresponding data are taken from the experimental work of \citet{delville2010transmission}. At the same time, cubic elastic anisotropy is assumed for both austenite and martensite phases, with the elastic constants of austenitic NiTi ($c_{11}=162$, $c_{44}=34$, $c_{12}=129$, all in GPa) \citep{mercier1980single}, regardless of the Pd content. This choice is dictated by the fact that no reliable experimental data are available on the elastic constants of the NiTiPd austenite and martensite phases for different Pd contents. 

The following model parameters are used in all the simulations. Interfacial energy densities are selected as $\gamma_{0i}=\gamma_\text{am}=0.2$ J/m$^2$ for the austenite--martensite interfaces and $\gamma_{ij}=\gamma_\text{mm}=0.02$ J/m$^2$ for the martensite--martensite interfaces, following earlier studies \citep{petryk2010interfacial}. The chemical energy of the austenite phase $F_0^0=F_\text{a}=0$ and martensite phases $F_i^0=F_\text{m}^0=5$ MPa are adopted. This set of chemical energies is equivalent to a (uniform) temperature in the temperature range of pseudoelasticity, implying that the austenite phase is stable in stress-free conditions. 
The mobility parameters are set as $m_i=m=0.1$ (MPa s)$^{-1}$. Note that, since the mobility parameter $m$ is the only parameter in the model that involves the unit of time, only the ratio between $m$ and the indentation speed $v$ is a relevant parameter that characterizes the rate effects (this means that any combination of $v$ and $m$ that yields the same ratio would lead to the same simulation results).
Finally, the interface thickness parameter is assumed as $\ell=1.5$ nm, which corresponds to the theoretical interface thickness of $\lambda=\pi \ell=4.7$ nm, and thus an interface thickness-to-element size ratio of $\lambda/h \approx 2.5$. The parameter $\ell$ has been justifiably assumed following our preliminary analysis and in light of our prior modeling experience with phase-field modeling of martensitic transformation. Indeed, the ratio $\lambda/h$ together with the mesh density used are sufficient to capture the subtle microstructural features, while a finer mesh density does not qualitatively affect the simulation results, see \citep{tuuma2021phase} for the related discussion.

\section{Simulation results and discussion}\label{sec-sim}
A detailed discussion of the simulation results is reported in this section. The objective of the study in Section~\ref{sec-CuAlNi} is to thoroughly analyze the evolution of martensitic transformation in a CuAlNi single crystal during nanoindentation. In Section~\ref{sec-NiTiPd}, the study is performed for NiTiPd, and the role of the Pd content is investigated with the focus on kinematic compatibility and transformation hysteresis.
\subsection{CuAlNi single crystal under nanoindentation}\label{sec-CuAlNi}
Microstructure evolution in a [111]-oriented CuAlNi (i.e., with the [111] axis of the cubic austenite lattice aligned with the indentation direction) is investigated first. The analysis is then extended to study the impact of crystal orientation on the nanoindentation response. Finally, a parametric study is carried out with the aim to assess the role of important model features, specifically, elastic anisotropy and finite-strain kinematics, on the simulation results.  

\subsubsection{Microstructure evolution in a [111]-oriented CuAlNi}\label{sec-111}
\begin{figure}
\centering
\hspace*{-0.75cm}
\includegraphics[width=1.05\textwidth]{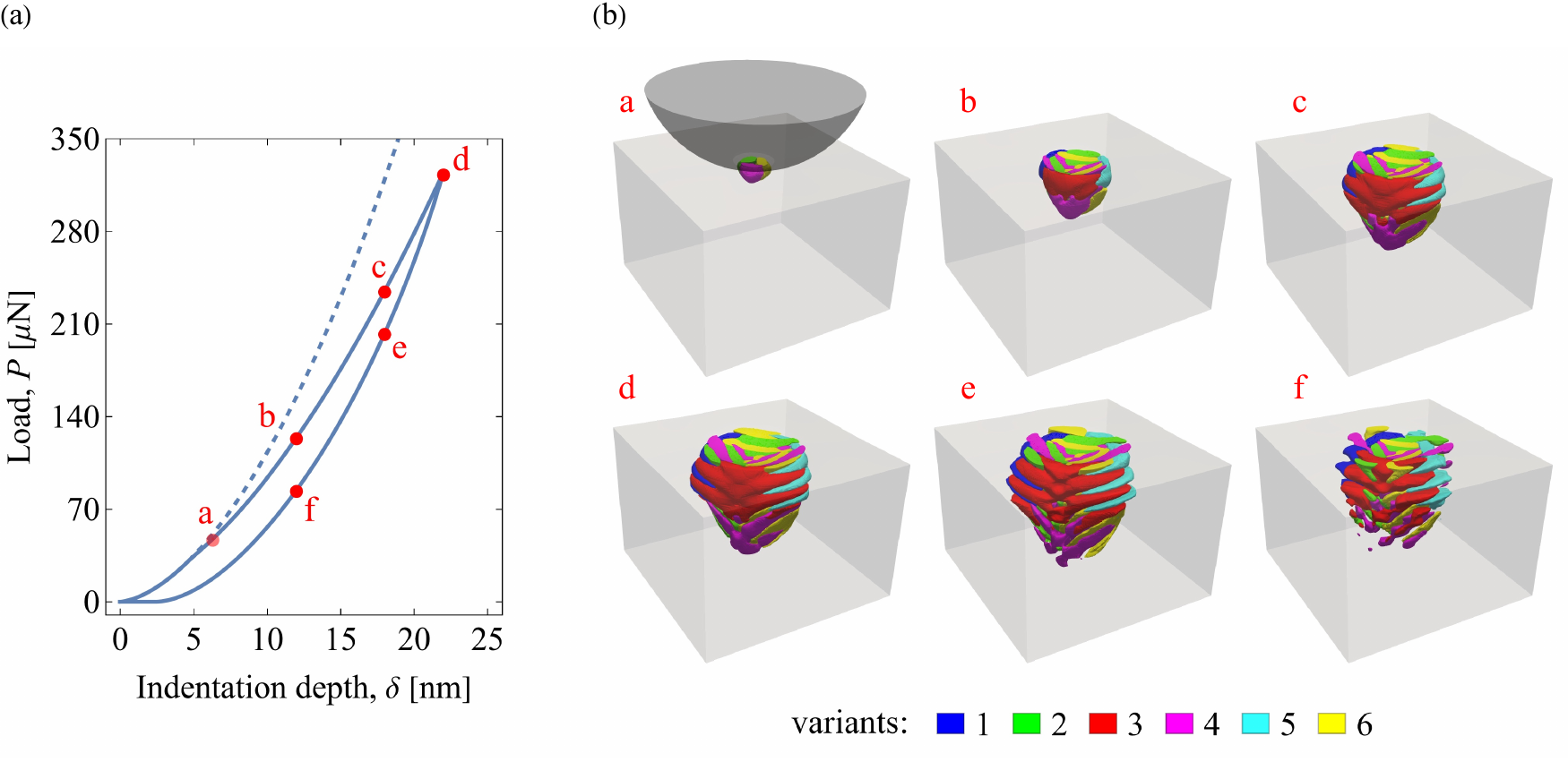}
\caption{The load--indentation depth ($P$--$\delta$) response (a) and the snapshots of the microstructure evolution (b) in a $[111]$-oriented CuAlNi single crystal. The dashed curve in panel (a) represents the elastic response. In panel (b), the austenite is shown as transparent, while each martensite variant is identified by a distinct color and the respective domain is represented by the volume fraction $\eta_i \geq 0.5$.}
\label{Fig-microFull}
\end{figure}

Fig.~\ref{Fig-microFull} depicts the mechanical response in terms of the indentation load $P$ versus the indentation depth $\delta$ together with representative snapshots of the microstructure evolution. The domain occupied by each martensite variant is identified by a distinct color and represents the spatial distribution of the corresponding order parameter $\eta_i$ (trimmed to $\eta_i \geq 0.5$). The domain occupied by the austenite is transparent.

The transformation initiates at the indentation depth of $\delta \approx 5$ nm. The nucleation event occurs in two stages, first, the nucleation of martensite variants 2, 4 and 6, which constitute the kernel of the transformed domain (see snapshot `a' in Fig.~\ref{Fig-microFull}(b)), followed by the nucleation of the remaining variants surrounding the kernel at the indentation depth of $\delta\approx7$ nm. The nucleation event is not accompanied by rapid changes in the $P$--$\delta$ response, rather, a gradual transition between the elastic and transformation branches is observed, in agreement with the nanoindentation experiments on SMA single crystals, e.g., \citep{yan2013anomalous,kumar2020assessment}. As the load increases, the microstructure develops more complex patterns, in particular, twin laminates are formed between various pairs of martensite variants, e.g., the pairs (1,3), (2,4) and (3,6). The twin laminates are clearly visible from the internal view of the microstructure, and the orientation of the twin planes agrees with the crystallographic theory of martensite \citep{Bhattacharya2003}. These aspects are discussed more in Section~\ref{sec-orientation}.

The transformed domain continues to grow slightly at the early stage of unloading (compare snapshots `d' and `e' in Fig.~\ref{Fig-microFull}(b)). This is a consequence of the viscous evolution law resulting from the viscous-type dissipation potential \eqref{Eq-diss}. In view of this, after unloading starts, the driving forces for the interface propagation do not change their sign immediately, hence the interfaces continue to propagate in the prior direction for some time, until the driving forces decrease to zero and change their sign. A similar effect is observed in atomistic simulations of martensite reorientation in SMAs \citep{wang2021molecular}. Thereafter, the reverse transformation commences via the shrinkage of the martensite domains. When the indentation depth is at $\delta \approx 3$ nm during unloading, the indenter separates from the top surface (thus $P=0$), while a remnant microstructure is still present (the corresponding microstructure is not shown here), and ultimately annihilates as time proceeds further. During the whole loading--unloading process, the microstructure exhibits an overall three-fold symmetry appearance, see Fig.~\ref{Fig-FDC2O}(b), originating from the [111] orientation of the crystal, though the local microstructural features do not strictly fulfill the symmetry condition.

\begin{figure}
\centering
\hspace*{0cm}
\includegraphics[width=0.86\textwidth]{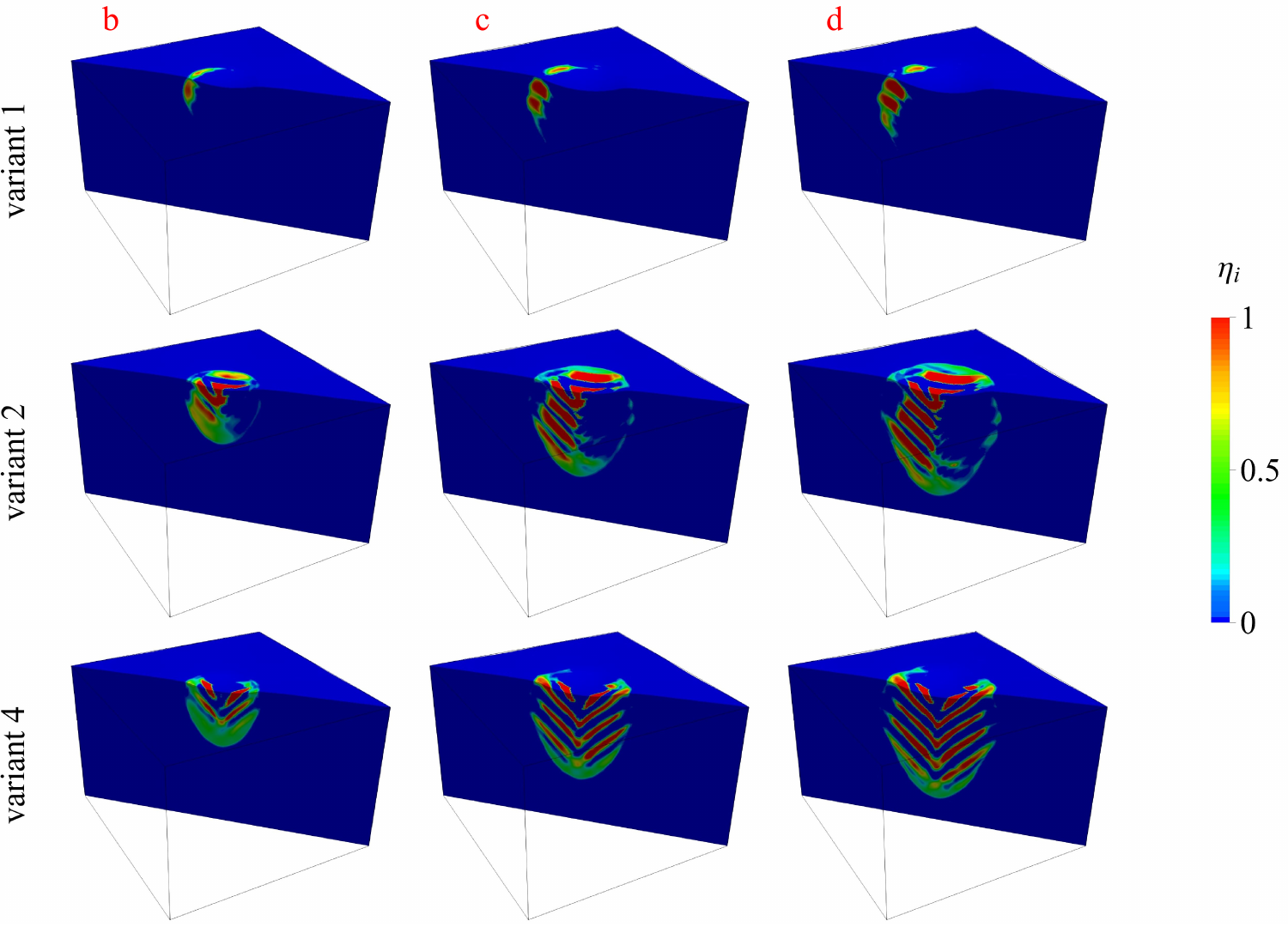}
\caption{Snapshots of the transformed domains of variants 1, 2 and 4 represented by the respective fields of the order parameters $\eta_i$ in a [111]-oriented CuAlNi single crystal. Each column of snapshots corresponds to a specific indentation depth $\delta$ during loading, see the respective labels in Fig.~\ref{Fig-microFull}(a).}
\label{Fig-diff}
\end{figure}

The post-processing applied to show the microstructure in Fig.\ref{Fig-microFull} (i.e., trimming to $\eta_i \geq 0.5$ and a transparent domain of the austenite) allows a better visualization of the microstructure and discloses the interaction between the domains of martensite variants. Nevertheless, such visualization does not reflect the diffuse character of the resulting microstructure. Fig.~\ref{Fig-diff} presents snapshots of the individual transformed domains of selected martensite variants (1, 2 and 4) represented by the actual field of the respective order parameters $\eta_i$. It follows that the microstructure is more diffuse at the bottom of the transformed domain. The more diffuse region is visible already at the early stage of transformation (see snapshot `b' in Fig.~\ref{Fig-diff}), indicating that it does not result from the interaction of the transformed domain with the bottom boundary of the computational domain. According to our experience \citep{tuuma2021phase}, a sufficiently small interface thickness parameter $\ell$ (and thus a finer mesh) would be needed to avoid such overly diffuse regions.

To understand the mechanism behind the variant selection, a simplified analysis based on the Schmid factor corresponding to uniaxial compression can be performed \citep{shield1995orientation,zhang2000variant}. The transformation Schmid factor is calculated as
\begin{equation}\label{Eq-schmid}
m_i=(\bfm{U}^\text{t}_i-\bfm{I}) \cdot \bar{\bm{\Sigma}}, \qquad i=1,\dots,6,
\end{equation}
where $\bfm{U}_i^\text{t}$ is the transformation stretch tensor of variant $i$, see Eq.~\eqref{Eq-stretch}, $\bfm{I}$ is the identity tensor, so that $\bfm{U}_i^\text{t}-\bfm{I}$ represents a strain measure, and $\bar{\bm{\Sigma}}=-\bfm{t} \otimes \bfm{t}$ characterizes the stress state (uniaxial compression), with $\bfm{t}$ denoting the unit vector specifying the crystal axis along which the indentation is applied. The calculation results are presented in Table~\ref{Tab-schmid}, which also includes the results corresponding to other crystal orientations, see the related discussion in Section~\ref{sec-orientation}. Table~\ref{Tab-schmid} suggests that the selection of the martensite variants can be reasonably predicted by this simple analysis. According to Table~\ref{Tab-schmid}, variants 2, 4 and 6, i.e., those with the largest Schmid factor $m_2=m_4=m_6=0.012$, are the favorable variants for the [111] crystal orientation. As shown in Fig.~\ref{Fig-microFull} and discussed above, these variants appear prior to the other variants and form the kernel of the transformed domain. On the other hand, the remaining variants (with negative Schmid factors $m_1=m_3=m_5=-0.014$) appear at a later stage of the evolution to accommodate the incompatibility caused by the developed microstructure.

\begin{table}
\caption{The transformation Schmid factors $m_i$ corresponding to uniaxial compression, see Eq.~\eqref{Eq-schmid}, calculated for all martensite variants and four crystal orientations. The largest Schmid factors for each orientation are shaded.}
\label{Tab-schmid}
\vspace{1ex}
\centering
\small{\begin{tabular}{ccccc}
\hline
 &[001]&[011]&[111]&[123] \\
\hline
$m_1$&$-0.043$&$\colorbox{shade}{0.020}$&$-0.014$&$-0.015$\\
$m_2$&$-0.043$&$\colorbox{shade}{0.020}$&$\colorbox{shade}{0.012}$&$\textcolor{white}{-}0.001$\\
$m_3$&$\colorbox{shade}{0.082}$&$\colorbox{shade}{0.020}$&$-0.014$&$\textcolor{white}{-}0.032$\\
$m_4$&$\colorbox{shade}{0.082}$&$\colorbox{shade}{0.020}$&$\colorbox{shade}{0.012}$&$\colorbox{shade}{0.043}$\\
$m_5$&$-0.043$&$-0.062$&$-0.014$&$-0.050$\\
$m_6$&$-0.043$&$-0.023$&$\colorbox{shade}{0.012}$&$-0.017$\\
\hline
\end{tabular}
}
\end{table}

\begin{figure}
\centering
\hspace*{-1.8cm}
\includegraphics[width=1.25\textwidth]{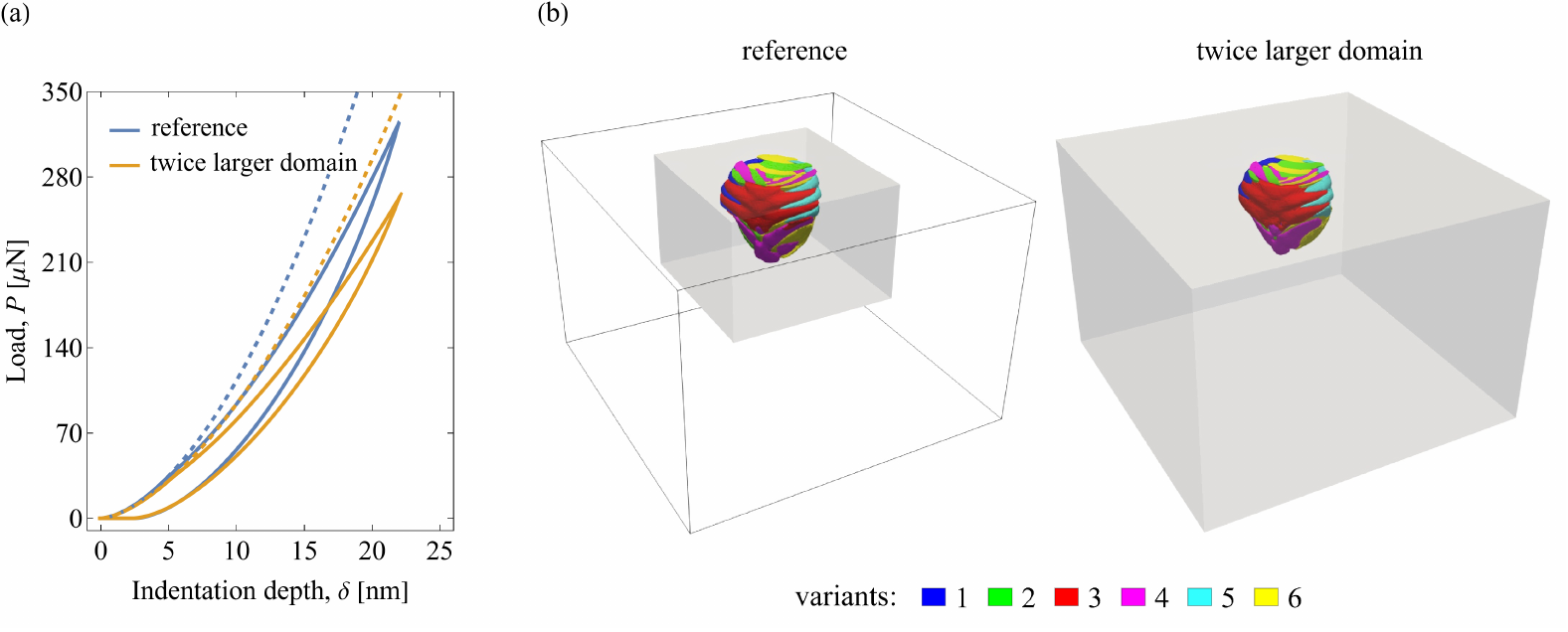}
\caption{Effect of the size of the computational domain: the load--indentation depth ($P$--$\delta$) response (a) and the microstructure at the maximum indentation depth $\delta_\text{max}=22$ nm (b) are compared for the reference case (with the computational domain of the size $400\times400\times280$ nm$^3$) and for a twice larger computational domain, both with the indenter radius $R=200$ nm. The dashed curves in panel (a) represent the elastic responses. The outer box in the left figure in panel (b) depicts the extent of the larger domain relative to the reference domain.}
\label{Fig-elbound}
\end{figure}

It can be seen in Fig.~\ref{Fig-microFull}(b) that the transformed domain occupies a significant part of the computational domain, and thus one may expect that the boundary conditions influence the microstructure evolution and the $P$--$\delta$ response. To clarify this issue, we have performed a simulation with a twice larger computational domain ($800\times800\times560$ nm$^3$), while all the other parameters, in particular, the indenter radius $R$ and the maximum indentation depth $\delta_\text{max}$, are kept unchanged. The element size $h$ is also kept unchanged in the central part where the microstructure develops, while it is significantly larger in the outer part, so that the problem involves approximately 100 million degrees of freedom. Fig.~\ref{Fig-elbound} compares the resulting $P$--$\delta$ response and the microstructure at the maximum indentation depth $\delta_\text{max}=22$~nm with those obtained from the reference simulation. It can be seen that, except a slight difference in the size of the transformed domain, no visible difference can be found in the microstructures (also concerning the internal microstructure). This is an important finding which confirms that the microstructure can be reliably predicted by using a relatively small computational domain. At the same time, the $P$--$\delta$ responses are markedly different, which is mainly due to the different interactions with the domain boundaries, as can be reckoned by comparing the corresponding elastic responses, see the dashed curves in Fig.~\ref{Fig-elbound}(a). This suggests that the $P$--$\delta$ response corresponding to a half-space (or a sufficiently large computational domain) could actually be estimated by adding the missing elastic deflection to the $P$--$\delta$ response of the reference simulation. This is not pursued here, as our main goal is the analysis of microstructure evolution.

\begin{figure}
\centering
\hspace*{0cm}
\includegraphics[width=0.96\textwidth]{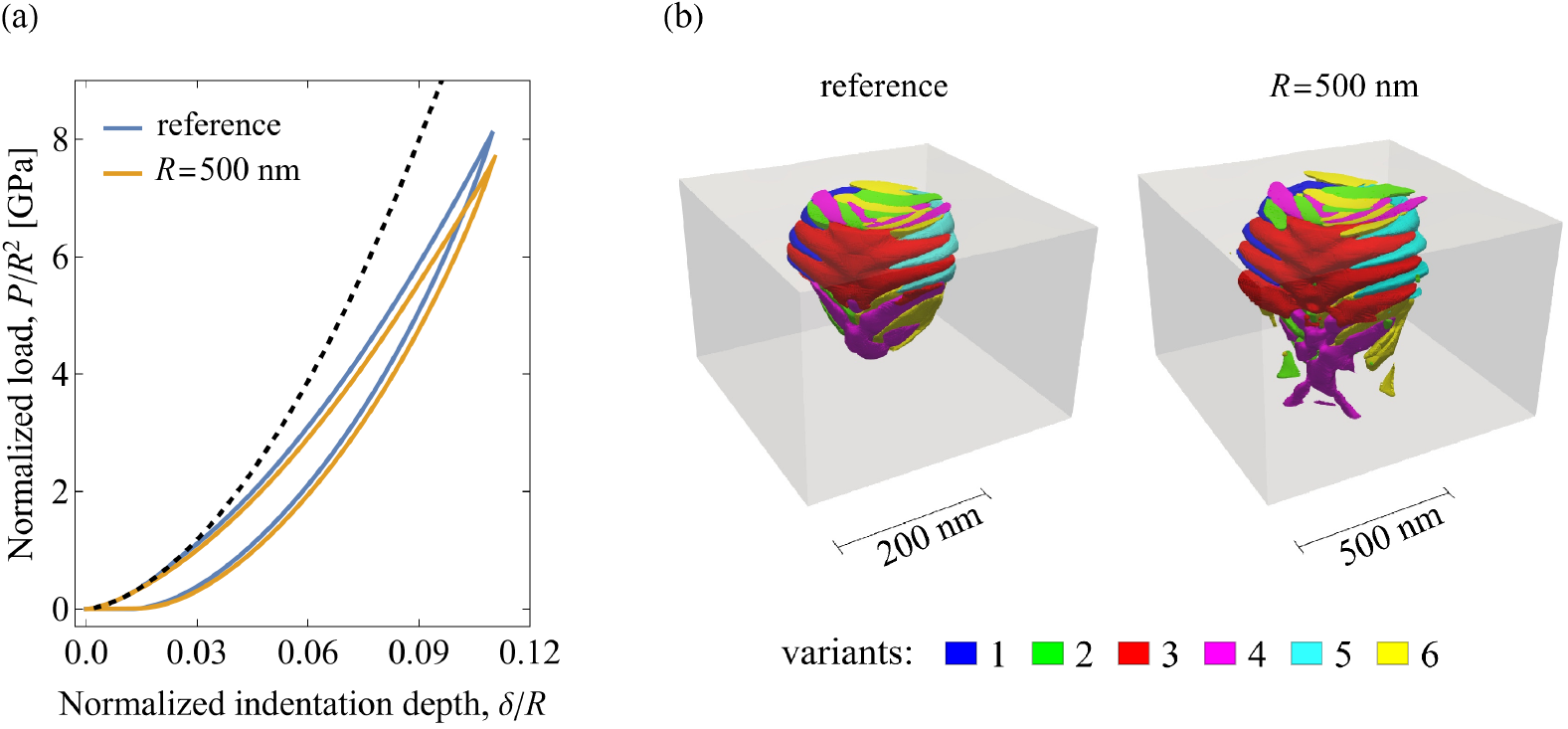}
\caption{Effect of the indenter radius $R$: the normalized load--indentation depth ($P/R^2$--$\delta/R$) response (a) and the microstructure at the maximum indentation depth (b) are compared for the reference case (with the computational domain of the size $400\times400\times280$ nm$^3$ and the indenter radius $R=200$ nm) and for the case with a 2.5 times larger computational domain ($1000\times1000\times700$ nm$^3$) and indenter radius ($R=500$ nm). Note that in the latter case, the element size $h$ and the interface thickness $\ell$ are also scaled by the factor of 2.5 with respect to the reference simulation. The dashed curve in panel (a) represents the normalized elastic response (identical in both cases).}
\label{Fig-Bigdomain}
\end{figure}

An additional simulation has also been performed in which all the dimensions are scaled by a factor of 2.5 with respect to the reference simulation (i.e., computational domain size is $1000\times1000\times700$ nm$^3$, indenter radius $R=500$ nm, element size $h=4.7$ nm, interface thickness parameter $\ell=3.75$ nm). To ensure the same effective mobility of the interfaces, the mobility parameter $m$ has been decreased 2.5 times, thus $m=0.04$ (MPa s)$^{-1}$ \citep{tuuma2021phase}. The results are reported in Fig.~\ref{Fig-Bigdomain}. It follows that the normalized $P$--$\delta$ response is very close to that of the reference simulation, and the prominent features of the microstructure are well reproduced. Extra domains of martensite variants, however, appear at the bottom surface. It has been verified that these extra domains emerge as a result of an overly diffuse microstructure at the bottom surface, and indicate that the interface thickness parameter $\ell=3.75$ nm is exceedingly large. Nevertheless, the results show that the microstructural features predicted for the relatively small indenter radius $R=200$ nm are representative also for larger indenter radii.

\subsubsection{Effect of crystal orientation}\label{sec-orientation}
\begin{figure}
\centering
\hspace*{-2.6cm}
\includegraphics[width=1.35\textwidth]{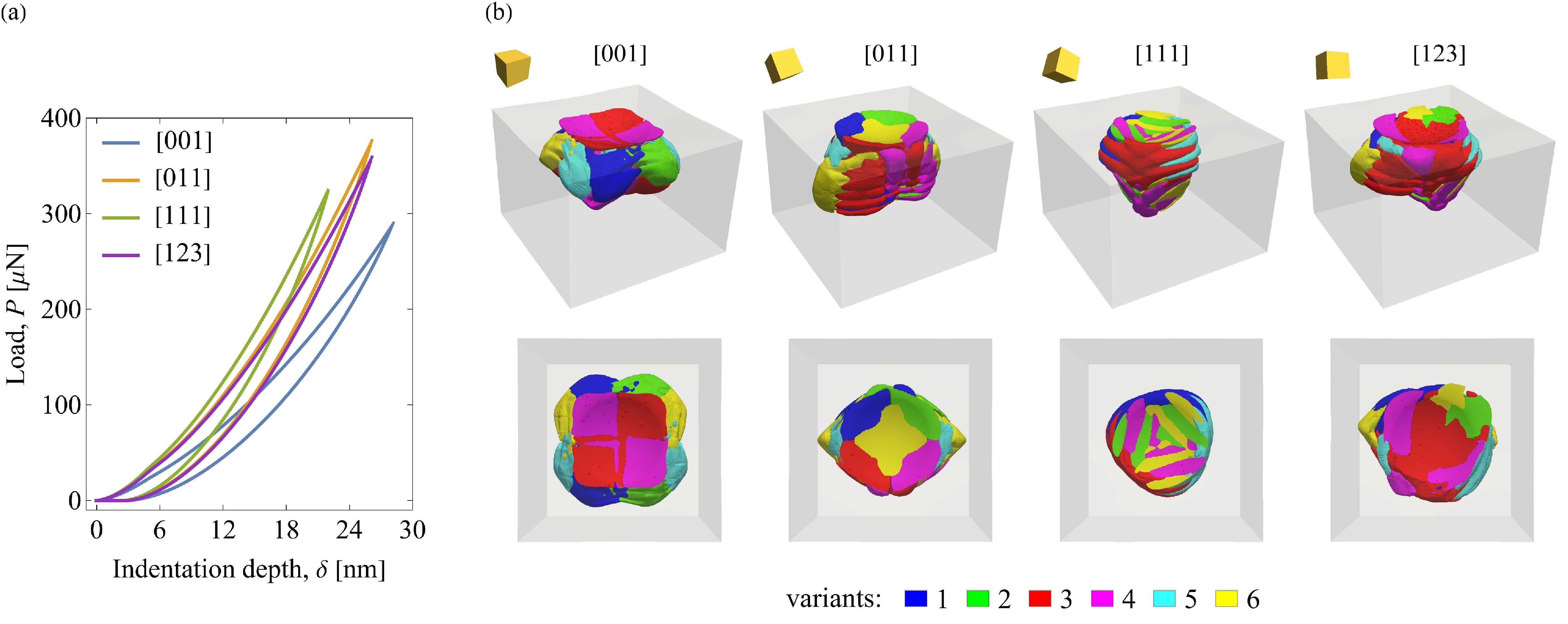}
\caption{Effect of crystal orientation: the load--indentation depth ($P$--$\delta$) response (a) and the microstructure at the maximum indentation depth $\delta_\text{max}$ (b) in CuAlNi. Note the different $\delta_\text{max}$ for different orientations. The orientation of the small cube in the top row of panel (b) indicates the orientation of the cubic lattice of the austenite.}
\label{Fig-FDC2O}
\end{figure}

In addition to the [111]-oriented CuAlNi reported in the previous section, simulations have been performed for three additional crystal orientations, namely [001] and [011], and a non-principal [123] orientation with no crystallographic symmetry. In Fig.~\ref{Fig-FDC2O}, the $P$--$\delta$ responses and the microstructures at the maximum indentation depth $\delta_\text{max}$ are compared for the four different orientations. The maximum indentation depth $\delta_\text{max}$ is set such that the transformed domain does not directly interact with the boundaries of the computational domain. Accordingly, in view of the distinct microstructure patterns evolved, $\delta_\text{max}$ is set differently for each orientation. A quick glance at the results in Fig.~\ref{Fig-FDC2O} reveals that the $P$--$\delta$ responses and the microstructures obtained are strongly influenced by the crystal orientation. A detailed analysis of the results is presented below. 

As shown in Fig.~\ref{Fig-FDC2O}(a), the $P$--$\delta$ responses obtained for [001] and [111] orientations exhibit the most compliant and the stiffest behaviours, respectively, while those of [011] and [123] are in between and are quite close to each other. The analysis of the $P$--$\delta$ responses is complemented by examining some contact-related quantities, in particular, the indentation hardness, which is calculated as
\begin{equation}
H=P/A_\text{p},
\end{equation}
with $P$ denoting the load and $A_\text{p}$ the projected contact area, see Table~\ref{Tab-contact}. To enable a meaningful comparison, the data provided in Table~\ref{Tab-contact} relate to the same indentation depth $\delta=22$ nm, with $P$ and $A_\text{p}$ being the corresponding instantaneous values. It follows that the maximum hardness $H=16.9$ GPa belongs to the [111] orientation, and the minimum hardness $H=10.9$ GPa to the [001] orientation. 
An important point that needs to be highlighted is that the values of the hardness calculated here are larger than those typically reported in the nanoindentation experiments of SMAs, e.g., \citep{yan2013anomalous,laplanche2014orientation}. This is attributed to the small size of the transformed domain (related to the limited size of the computational domain) and, thereby, the decisive role of the interfacial energy contribution. As a result, a relatively high indentation load is required to induce the microstructure, see \citep{rezaee2020phase} for the related discussion. The overprediction of the hardness is also due to the effect of boundary conditions, as illustrated in Fig.~\ref{Fig-elbound}. Finally, we note that the stress-induced martensitic transformation is here the only inelastic deformation mechanism, while plasticity (dislocation slip), even if limited, may contribute to the lower hardness observed in the experiments. Actually, at very small scales, activity of plastic slip may be limited due to the strengthening effect of dislocation starvation \citep{greer2005size} and strain gradients \citep{fleck1994strain}. In fact, a (nearly) complete deformation recovery has been reported in the nanoindentation experiments on NiTi \citep{amini2013phase}, while, depending on the indenter radius, both complete and incomplete recovery have been reported in \citep{kumar2020assessment}. Inclusion of plasticity would be a natural next step to enhance the model, e.g., \citep{yeddu2012three,xu2020phase,rezaee2022deformation}, but it is beyond the scope of this work.

Another quantity included in Table~\ref{Tab-contact} is the ratio between the projected contact area $A_\text{p}$ and the nominal contact area $A_\text{n}$, the latter determined by simple geometry as $A_\text{n}=\pi \delta(2R-\delta)$, and thus $A_\text{n}=26.1\times10^3$ nm$^2$ for $\delta=22$ nm. In the context of plasticity, the $A_\text{p}/A_\text{n}$ ratio, known as $c^2$, is related to the amount of pile-up/sink-in and characterizes the degree of strain hardening, e.g., \citep{hill1989theoretical,petryk2017direct}. Table~\ref{Tab-contact} indicates that the $A_\text{p}/A_\text{n}$ ratio is weakly influenced by the crystal orientation, however, analogous to the hardness $H$, it shows the largest and smallest values for [111] and [001] orientations, respectively. 

We stress again that we are not aware of any experiments on CuAlNi single crystal that could be used to verify quantitatively the predicted mechanical response for different crystal orientations.

\begin{table}
\caption{Contact-related quantities calculated for different crystal orientations. The values correspond to the same indentation depth $\delta$=22 nm, i.e., the maximum indentation depth for [111]-oriented CuAlNi, see Fig.~\ref{Fig-FDC2O}(a).}
\label{Tab-contact}
\vspace{1ex}
\centering
\small{\begin{tabular}{lcccc}
\hline
 &[001]&[011]&[111]&[123] \\
\hline
Indentation load, $P$ [$\mu$N]&195&285&324&273 \\
Projected contact area, $A_\text{p}$ [$10^3$ nm$^2$]&17.9&18.3&19.1&18.8 \\
$A_\text{p}/A_\text{n}$ ratio, & 0.69&0.70&0.73&0.72 \\
Hardness, $H$ [GPa]&10.9&15.6&16.9&14.5 \\
\hline
\end{tabular}
}
\end{table}

\begin{figure}
\centering
\hspace*{-1.4cm}
\includegraphics[width=1.2\textwidth]{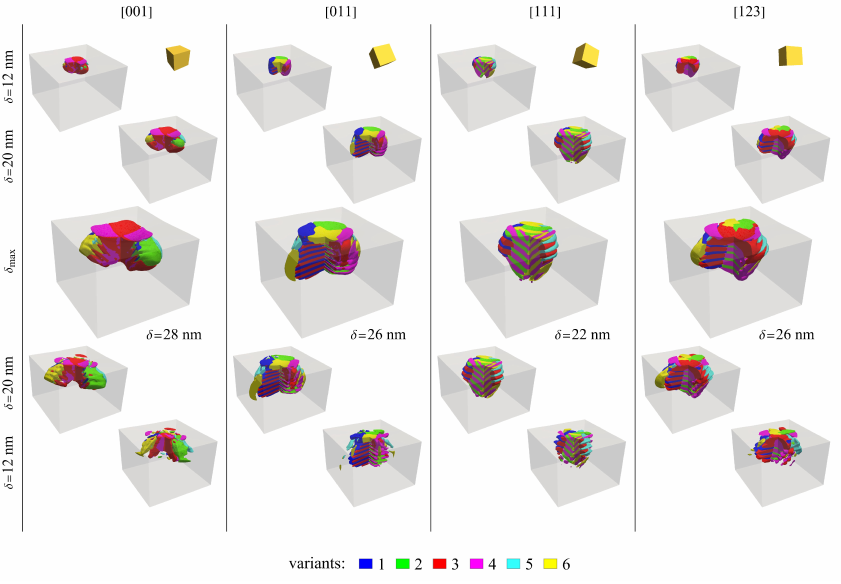}
\caption{Microstructure evolution (during loading and unloading) in CuAlNi for different crystal orientations. To display the internal features of the microstructure, one quarter of the transformed domain is removed. The orientation of the small cube in the top row indicates the orientation of the cubic lattice of the austenite.}
\label{Fig-C2OAll}
\end{figure}

Next, we elaborate more on the impact of the crystal orientation on the microstructure. Fig.~\ref{Fig-FDC2O}(b) clearly shows the diversity in the microstructure patterns evolved for different orientations. Despite the presence of all variants for all the orientations, the general shape of the transformed domain, the arrangement of the martensite variants, and their interaction with each other is markedly affected by the orientation. It is visible from the top view that the transformed domain for [001], [011] and [111] orientations possesses an overall four-, two- and three-fold symmetry, respectively. To investigate the microstructures in more detail, representative snapshots of the microstructure evolution for different orientations are illustrated in Fig.~\ref{Fig-C2OAll}. The internal features of the microstructures are displayed by removing one quarter segment of the transformed domain. 
Fig.~\ref{Fig-C2OAll} reveals that, for each crystal orientation, a particular set of martensite variants constitute the kernel of the transformed domain. In most cases, the generated variants are correctly predicted by the Schmid analysis (those with the highest Schmid factors, see Table~\ref{Tab-schmid}).

From the internal view of the microstructures in Fig.~\ref{Fig-C2OAll}, twin laminates can be observed at several places. The laminates comprise layers of two martensite variants separated by (approximately) parallel interfaces, i.e., twin planes. Although the formation of the laminates does not disrupt the overall symmetry of the microstructure, it breaks the local symmetry between the variants involved in twinning. This, in particular, can be seen for the [011] and [111] microstructures with fully-developed twin laminates.

\begin{figure}
\centering
\includegraphics[width=0.82\textwidth]{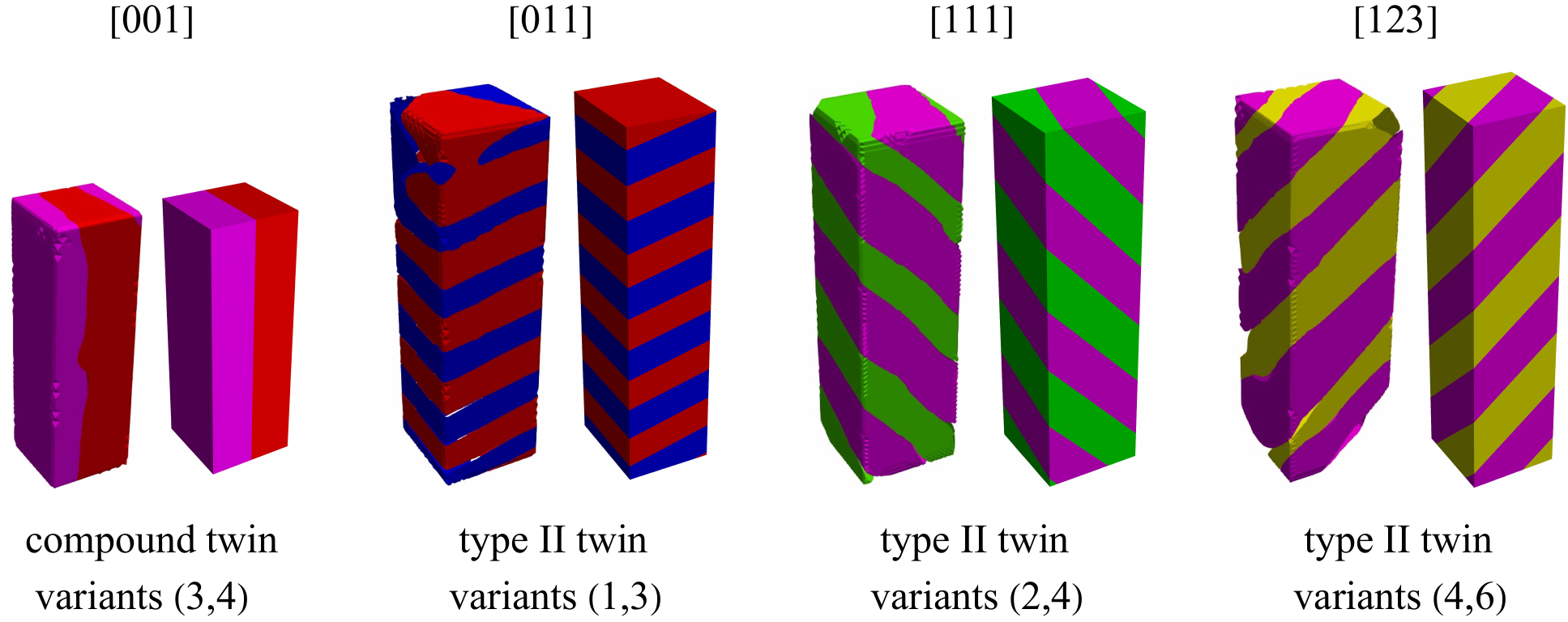}
\caption{Comparison between the orientation of the twin planes resulting from the simulations (left images, shown in the reference configuration) and those predicted by the crystallographic theory (right images, with the twin spacing adjusted manually). The twin laminates have been extracted from the respective microstructures at the maximum indentation depth $\delta_\text{max}$, see Fig.~\ref{Fig-C2OAll}.}
\label{Fig-twinning}
\end{figure}

Upon identification of the twin plane orientations, it has been found out that in all cases, the developed twins are either type II or compound twins, while no twin laminates with type I twins have been detected. This finding is consistent with the experiments on CuAlNi, which confirmed the preference for type II and compound twins under compression \citep{novak2006transformation,ge2022deformation}. Recall that in the present model the interfacial energy $F_\text{int}$, see Eq.~\eqref{Eq-int}, is isotropic, and thus does not distinguish between twins of different types. Fig.~\ref{Fig-twinning} shows representative twin laminates extracted from the individual microstructures and compared to the prediction by crystallographic theory \citep{Bhattacharya2003}. In the simulated microstructures, the characteristic twin spacing results from the interplay between the interfacial energy and the elastic microstrain energy \citep{petryk2010interfacial,tuuma2016size}. The spacing in the images illustrating the predictions by crystallographic theory has been adjusted manually. 

\subsubsection{Effect of elastic anisotropy}\label{sec-elastic}
The goal of this section is to highlight the impact of elastic anisotropy on the simulation results. In the previous section, we addressed the problem of microstructure evolution for different crystal orientations, where the elastic anisotropy of cubic symmetry for the austenite phase and of orthorhombic symmetry for the martensite phases have been accounted for (this case is referred to as `full anisotropy' in the sequel). The same problem is here approached considering two additional cases, i.e., the case with the cubic elastic anisotropy for both austenite and martensite phases, with the elastic constants of the austenitic CuAlNi (referred to as `homogeneous cubic' in the sequel), and the case with isotropic elasticity. In the latter case, the corresponding shear and bulk moduli ($\mu=39$ GPa and $\kappa=128$ GPa) are obtained by applying the Voigt-Reuss-Hill averaging \citep{hill1952elastic} to the elastic constants of austenitic CuAlNi.

\begin{figure}
\centering
\hspace*{-0.4cm}
\includegraphics[width=1.05\textwidth]{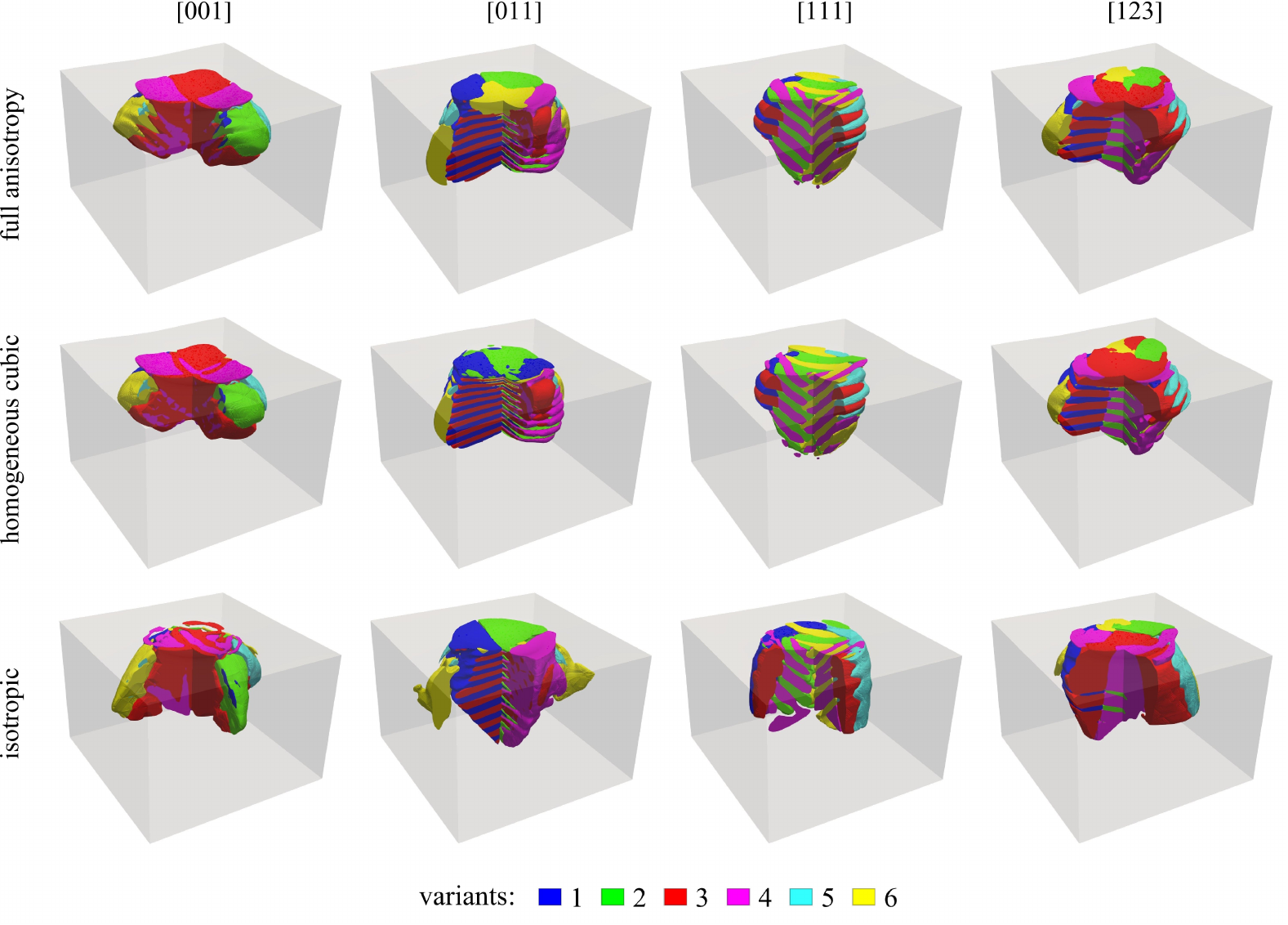}
\caption{The effect of elastic anisotropy on the microstructure in CuAlNi. The snapshots are taken at the maximum indentation depth $\delta_\text{max}$ (different for each orientation, see Fig.~\ref{Fig-FDC2O}). In the reference case of full anisotropy, distinct elastic anisotropy of all phases is accounted for. In the homogeneous cubic case, cubic anisotropy of the austenite is adopted for all phases. In the isotropic case, all phases have the same isotropic elastic properties.}
\label{Fig-elasticityFull}
\end{figure}

The effect of elastic anisotropy on the microstructure is depicted in Fig.~\ref{Fig-elasticityFull}, where the snapshots are taken at the maximum indentation depth $\delta_\text{max}$, as in Fig.~\ref{Fig-FDC2O}. From the comparison of the microstructures between the full anisotropy and the homogeneous cubic cases, it can be seen that the overall appearance of the microstructures is quite similar. Some small differences, however, are noticeable. This, in particular, concerns the microstructures for the [011] orientation, for instance, the absence of the martensite variant 6 within the kernel of the transformed domain in the homogeneous cubic case, or the difference in the twin spacing (e.g., the average twin spacing is about 22 nm in the full anisotropy case, while it is about 19 nm in the homogeneous cubic case). At the same time, it is immediate to notice significant microstructural differences between the isotropic case and the other two cases. The differences are mainly concerned with the overall shape of the transformed domain, and the spatial arrangement of the martensite variants. 

Upon investigating the $P$--$\delta$ responses, it has been found out that almost for all orientations, the homogeneous cubic and isotropic cases exhibit the stiffest and the most compliant mechanical behaviour, respectively, where the difference is the maximum for the [111] orientation (about $40\%$ of the load $P$ at the maximum indentation depth $\delta_\text{max}=22$ nm). Moreover, in each elasticity case, the $P$--$\delta$ curves of different orientations exhibit the same ordering as those in full anisotropy, see Fig.~\ref{Fig-FDC2O}(a). The corresponding plots are not provided here for brevity.

\subsubsection{Effect of finite-strain kinematics}\label{sec-finitesmall}
In this section we demonstrate the crucial role of finite-strain kinematics on the results obtained. To this end, the model is reformulated in the small-strain framework and is employed to simulate the microstructure evolution for different crystal orientations. 
The material parameters are the same as those introduced in Section~\ref{sec-setup}, in particular, full elastic anisotropy is taken into account. 
The reformulation of the model from the finite-strain to the small-strain framework is straightforward, however, it is briefly outlined here for completeness. In the small-strain model, the total strain $\bm{\upvarepsilon}=\frac{1}{2}(\nabla \bfm{u}+(\nabla \bfm{u})^\text{T})$ admits an additive decomposition into the elastic part $\bm{\upvarepsilon}^\text{e}$ and transformation part $\bm{\upvarepsilon}^\text{t}$, i.e., $\bm{\upvarepsilon}=\bm{\upvarepsilon}^\text{e}+\bm{\upvarepsilon}^\text{t}$. The elastic strain energy is then expressed as a quadratic function of the elastic strain~$\bm{\upvarepsilon}^\text{e}=\bm{\upvarepsilon}-\bm{\upvarepsilon}^\text{t}$, cf.\ Eq.~\eqref{Eq-elasticEn},
\begin{equation}
F_\text{el}=\frac{1}{2}(\bm{\upvarepsilon}-\bm{\upvarepsilon}^\text{t}) \cdot \mathbb{L}(\bm{\upvarepsilon}-\bm{\upvarepsilon}^\text{t}), \quad \bm{\upvarepsilon}^\text{t}=\sum_{i=1}^N \eta_i \bm{\upvarepsilon}_i^\text{t},
\end{equation}
where $\bm{\upvarepsilon}_i^\text{t}$ denotes the transformation strain of the martensite variant $i$ and is defined as $\bm{\upvarepsilon}_i^\text{t}=\bfm{U}^\text{t}_i-\bfm{I}$, see Eq.~\eqref{Eq-stretch}. The other energy components remain the same as those in the finite-strain model. 

\begin{figure}
\centering
\hspace*{-0.4cm}
\includegraphics[width=1.05\textwidth]{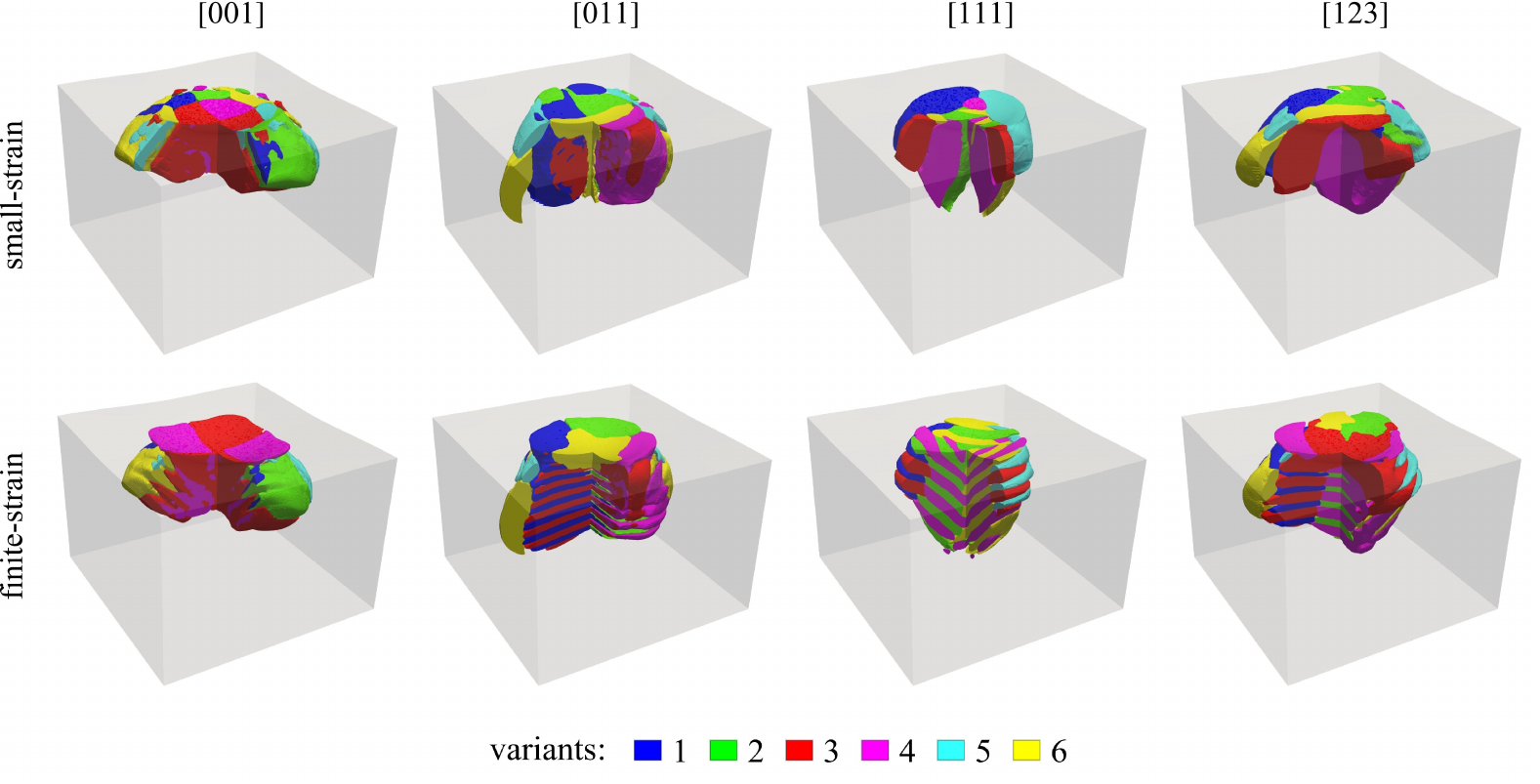}
\caption{The effect of finite-strain kinematics on the microstructure in CuAlNi: predictions of the small-strain theory (top row) are compared to the kinematically exact finite-strain theory (reference case, bottom row). The snapshots are taken at the maximum indentation depth $\delta_\text{max}$ (different for each orientation, see Fig.~\ref{Fig-FDC2O}).}
\label{Fig-smallfinite}
\end{figure}

Fig.~\ref{Fig-smallfinite} compares the microstructures predicted by the finite- and small-strain models for different crystal orientations. It follows that, indeed, the effect of the finite-strain kinematics is striking. Although some microstructural features, namely the overall symmetry of the microstructure and the formation of all martensite variants, are reproduced in the small-strain simulations, strong discrepancies are observed with respect to those of finite-strain simulations. Interestingly, no twin laminates have been formed in the small-strain simulations, instead, the microstructures have developed bulky martensite domains. 

The main difference between the small- and finite-strain theory is in the way the rotations are handled, and this may be the reason behind the significant difference between the microstructures predicted by the two theories, as discussed in detail by \citet{finel2010phase}. Actually, the lack of fine twin laminates in our small-strain simulations bears qualitative similarities to the results reported in \citep{finel2010phase} for a different problem, where microtwins disappear in the small-strain setting, but persist in the finite-strain setting. 

In addition to the microstructures, as expected, the $P$--$\delta$ responses are also affected by the choice of the deformation kinematics. The effect is quantitatively different depending on the crystal orientation, however, in all cases, it is less than $10\%$ of the load $P$ at the maximum indentation depth $\delta_\text{max}$ (the corresponding plots are not shown here for brevity).

\subsection{NiTiPd single crystal under nanoindentation: the role of the Pd content}\label{sec-NiTiPd}
In this section, we exploit the nanoindentation problem setup to study the role of the Pd content in the transformation behaviour of a NiTiPd single crystal. In this alloy, the Pd content can be adjusted such that a very low transformation hysteresis is achieved \citep{cui2006combinatorial,delville2010transmission}. According to the geometrically nonlinear crystallographic theory of martensite \citep{ball1992proposed,james2005way}, the following conditions have to be satisfied so that an SMA material exhibits a very low transformation hysteresis: (i) no transformation volume change, mathematically expressed as $J_\text{t}=\det(\bfm{U}_i^\text{t})=\lambda_1 \lambda_2 \lambda_3=1$, with $\lambda_1 \leq \lambda_2 \leq \lambda_3$ as the ordered eigenvalues of the transformation stretch tensors $\bfm{U}_i^\text{t}$, and (ii) phase compatibility\footnote{The cofactor conditions are yet another set of compatibility conditions (called supercompatibility) that if satisfied lead to an extremely low transformation hysteresis \citep{chen2013study,gu2018phase}. However, these conditions are out of the scope of the current study and will not be considered further on.}, expressed as $\lambda_2=1$, which implies that a compatible planar interface can be formed between austenite and a single variant of martensite. It has been observed in the experiment on ternary Ni-based SMAs under temperature-induced phase transformation \citep{cui2006combinatorial} that there is only a weak correlation between the transformation 
volume change $J_\text{t}$ and thermal hysteresis. At the same time, the change in $\lambda_2$ significantly impacts the microstructure evolution (i.e., as $\lambda_2$ approaches unity, the material tends to develop twinless austenite--martensite microstructure) and, thereby, impacts the thermal hysteresis. 

\begin{figure}
\begin{center}
\hspace*{-1.5cm}\includegraphics[width=1.2\textwidth]{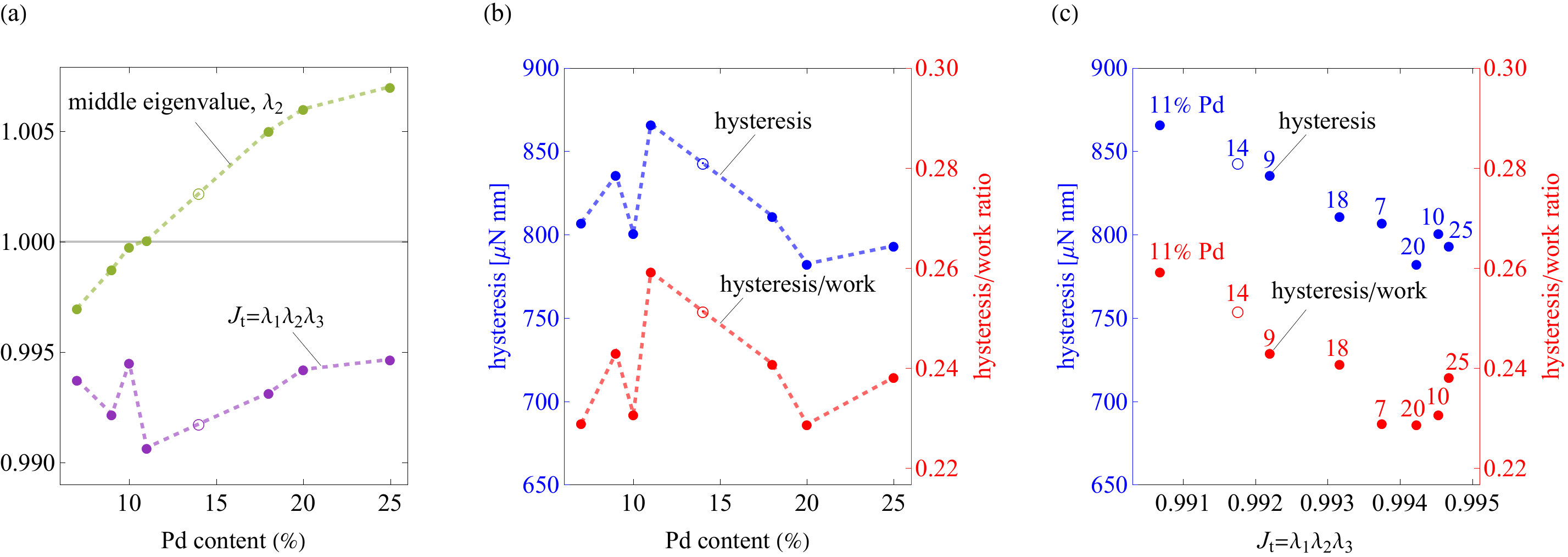}
\end{center}
\caption{(a) The graphs of the middle eigenvalue $\lambda_2$ and the transformation volume change $J_\text{t}=\lambda_1 \lambda_2 \lambda_3$ as a function of the Pd content, with the corresponding crystallographic data taken from \citep{delville2010transmission}. (b,c) The graphs of the hysteresis and the hysteresis/work ratio as a function of (b) the Pd content and (c) the transformation volume change $J_\text{t}$. Note that, due to the lack of experimental data in the Pd range from 11\% to 18\%, an additional simulation has been performed for 14\% Pd, where the crystallographic lattice parameters of this hypothetical alloy are obtained by a linear interpolation. The corresponding points on the graphs are denoted by empty circles.}
\label{Fig-NiTiPdAnal}
\end{figure}

Using the crystallographic data reported by \citet{delville2010transmission}, see Table~1 therein, we provide in Table~\ref{Tab-schmidPd} the transformation stretch parameters ($\alpha$, $\beta$, $\gamma$) for different Pd contents. The stretch parameters are then used to plot the graphs of the transformation volume change $J_\text{t}$ and the middle eigenvalue $\lambda_2$ as a function of the Pd content, see Fig.~\ref{Fig-NiTiPdAnal}(a). It follows that the middle eigenvalue $\lambda_2$ is closest to unity for the Pd content of 11\%, while $\lambda_2>1$ and $\lambda_2<1$ for the Pd content, respectively, greater and lower than 11\%. At the same time, $J_\text{t}$ has the largest deviation from unity for the Pd content of 11\%, and hence the largest transformation volume change. 
Having these characteristics and the above experimental findings in mind, the goal we pursue in this study is to investigate whether and how the Pd content and the two low-hysteresis indicators ($J_\text{t}$ and $\lambda_2$) impact the nanoindentation response of NiTiPd. 

\begin{table}
\caption{The transformation stretch parameters ($\alpha$, $\beta$, $\gamma$) and the largest transformation Schmid factors, see Eq.~\eqref{Eq-schmid}, corresponding to uniaxial tension (superscript `tens') and uniaxial compression (superscript `comp') calculated for NiTiPd with different Pd contents. The crystallographic data for different Pd contents have been taken from the work of \citet{delville2010transmission}. Note that the Pd content does not influence the sets of favoured martensite variants with the largest Schmid factors.}
\label{Tab-schmidPd}
\vspace{1ex}
\centering
\small{\begin{tabular}{cccccccc}
\hline
 &$7\%$ Pd&$9\%$&$10\%$&$11\%$&$18\%$&$20\%$&$25\%$ \\
\hline
$\alpha$ &0.9970&0.9988&0.9998&1.0001&1.0050&1.0060&1.0070 \\
$\beta$ &0.9398&0.9341&0.9332&0.9280&0.9227&0.9244&0.9167 \\
$\gamma$ &1.0606&1.0635&1.0659&1.0674&1.0710&1.0691&1.0775 \\
\hline
$m_\text{[001]}^\text{tens}$&0.0288&0.0311&0.0329&0.0338&0.0380&0.0376&0.0423\\
$m_\text{[011]}^\text{tens}$&0.0606&0.0635&0.0659&0.0674&0.0710&0.0691&0.0775\\
$m_\text{[111]}^\text{tens}$&0.0203&0.0204&0.0217&0.0209&0.0216&0.0209&0.0239\\
$m_\text{[001]}^\text{comp}$&0.0602&0.0659&0.0668&0.0720&0.0773&0.0756&0.0833\\
$m_\text{[011]}^\text{comp}$&0.0157&0.0174&0.0170&0.0191&0.0197&0.0190&0.0205\\
$m_\text{[111]}^\text{comp}$&0.0221&0.0228&0.0224&0.0239&0.0224&0.0212&0.0231\\
\hline
\end{tabular}
}
\end{table}

\begin{figure}
\centering
\includegraphics[width=0.78\textwidth]{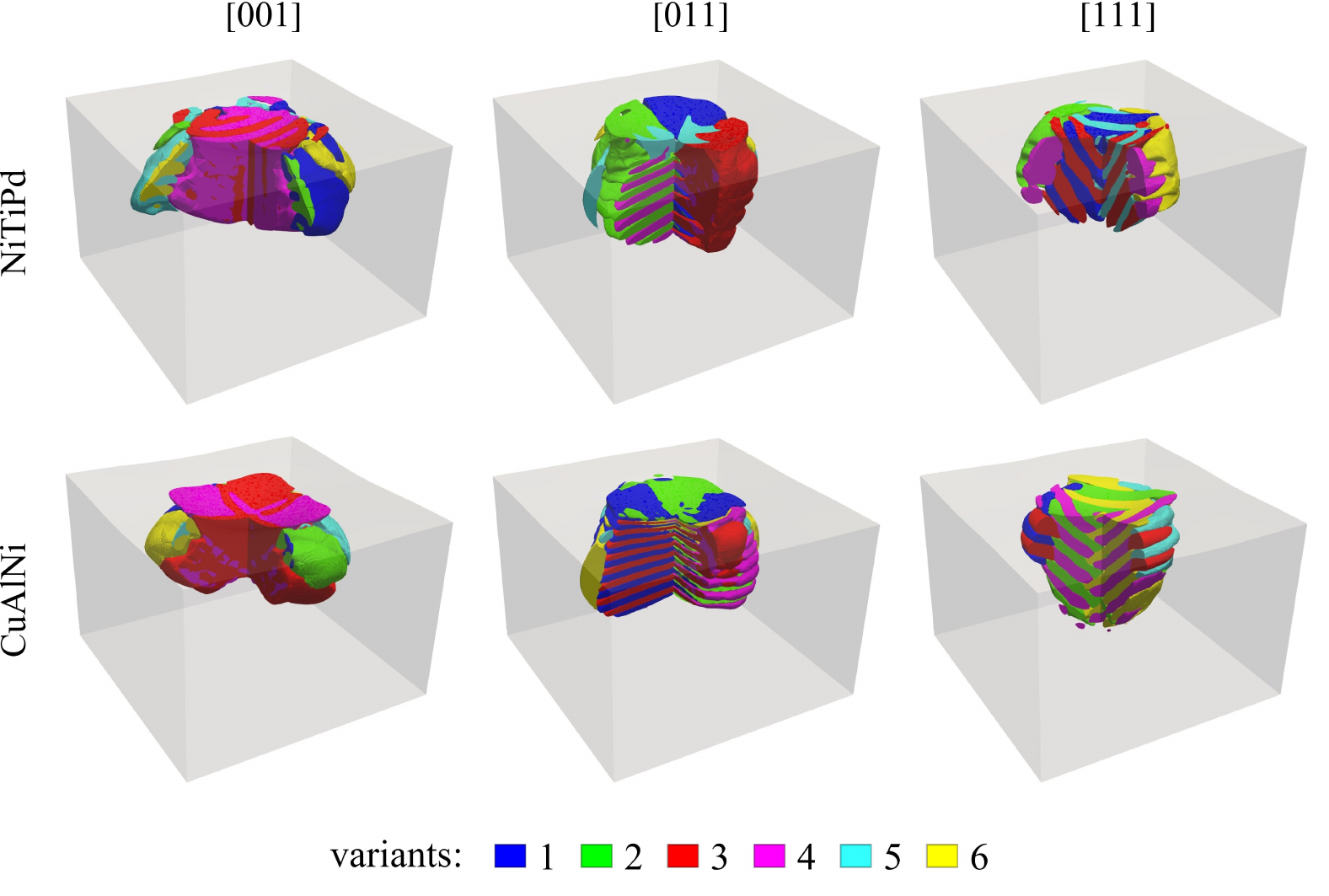}
\caption{The microstructures obtained for NiTiPd (with the Pd content of 11\%) and CuAlNi for different crystal orientations. For each orientation, the snapshots are taken at the same indentation depth. Both CuAlNi and NiTiPd alloys undergo a cubic-to-orthorhombic transformation, but the respective transformation stretch parameters ($\alpha$, $\beta$, $\gamma$) are different, which visibly influences the predicted microstructures.}
\label{Fig-NiTiPd-CuAlNi}
\end{figure}

\begin{figure}
\begin{center}
\hspace*{-2.0cm}\includegraphics[width=1.25\textwidth]{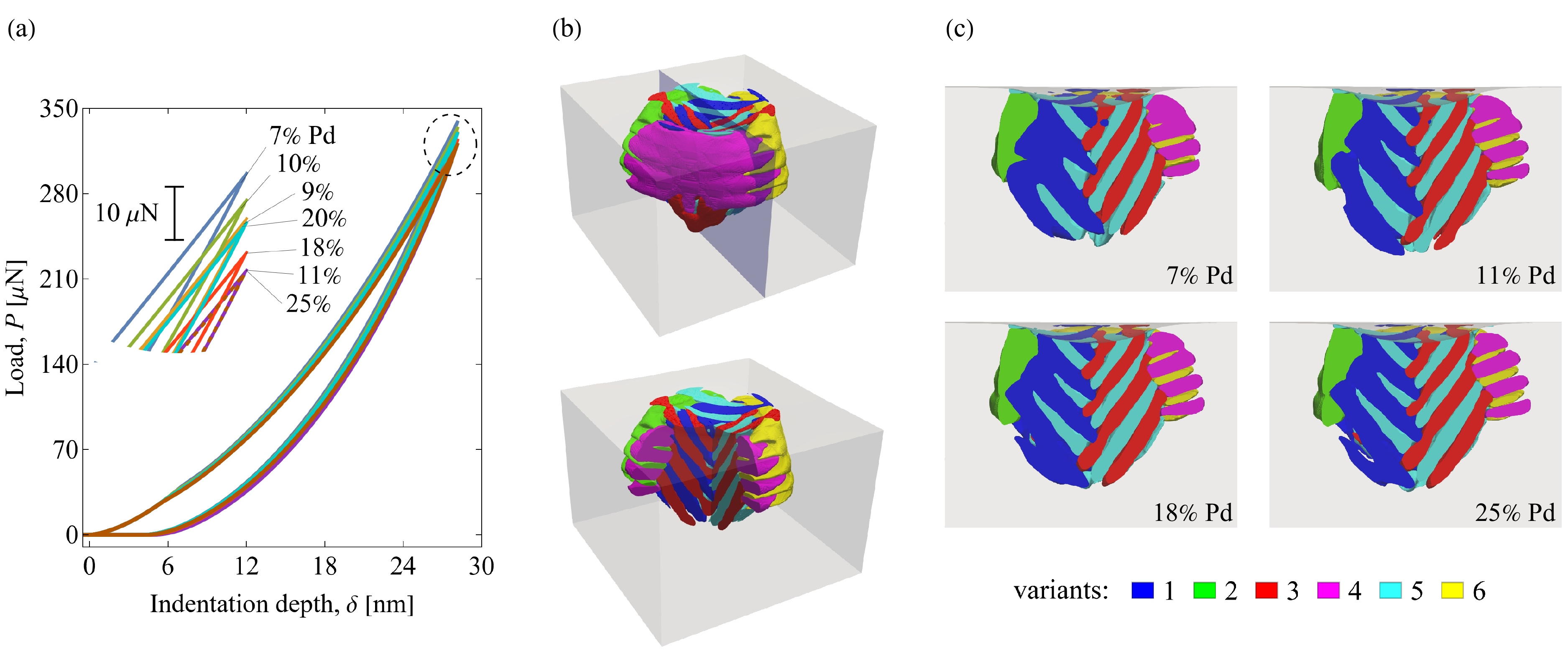}
\end{center}
\caption{(a) The effect of the Pd content on the load--indentation depth ($P$--$\delta$) response. (b) The overall view (top) and the internal features (bottom) of the microstructure for 11\% Pd at the maximum indentation depth $\delta_\text{max}=28$ nm. (c) The effect of the Pd content on the fine features of the microstructure. The microstructures in panel (c) are displayed through a vertical plane along the centerline of the computational domain (shown in panel (b)).}
\label{Fig-NiTiPdMicFD}
\end{figure}

As the starting point of this study, the microstructures obtained for NiTiPd are compared to those of CuAlNi for different crystal orientations. For this comparison, we have chosen the microstructures predicted for the Pd content of 11\%. Nevertheless, as discussed later and shown in Fig.~\ref{Fig-NiTiPdMicFD}(c), the effect of the Pd content on the microstructure evolution is negligible and the resulting microstructures for different Pd contents are substantially similar. For a more consistent comparison, we consider CuAlNi with homogeneous cubic elasticity, hence the same kind of elastic anisotropy is employed for both materials, see Section~\ref{sec-elastic}. Fig.~\ref{Fig-NiTiPd-CuAlNi} shows that, although CuAlNi and NiTiPd have similar microstructure patterns, they exhibit some significant differences, for instance, in the shape of the individual martensite domains or in the selection of the martensite variants in twin laminates. The differences obviously stem from two sources, namely the elasticity and the transformation stretches. Firstly, the two materials are characterized by different anisotropy (Zener) coefficients $A=2c_{44}/(c_{11}-c_{12})$, with $A=12$ (highly anisotropic) for CuAlNi and $A=2.1$ (weakly anisotropic) for NiTiPd. Secondly, the stretch parameters for NiTiPd lie in a different range than those of CuAlNi. In particular, the stretch parameter $\beta$ (which is equal to the second eigenvalue $\lambda_2$) in NiTiPd is close to unity, $\beta=1 \pm 0.0075$, while $\beta=0.9178$ in CuAlNi. The effect of crystal orientation on the load--indentation depth ($P$--$\delta$) response in NiTiPd is similar to that in CuAlNi shown in Fig.~\ref{Fig-C2OAll}(a), and thus it is omitted here for brevity.

We now focus on the role of the Pd content in the transformation behaviour of NiTiPd. Before delving into the main analysis, we examine the impact of the Pd content on the transformation Schmid factors. Table~\ref{Tab-schmidPd} reports the largest transformation Schmid factors (among those calculated for all the six martensite variants) corresponding to uniaxial tension, with $\bar{\bm{\Sigma}}=\bfm{t} \otimes \bfm{t}$, and uniaxial compression, with $\bar{\bm{\Sigma}}=-\bfm{t} \otimes \bfm{t}$, cf.~Eq.~\eqref{Eq-schmid}. For a given orientation and loading, the largest Schmid factor provides an estimate (in fact, an upper bound) of the maximum attainable transformation strain. It is seen that the Schmid factors calculated for the [001] orientation for compression (and also for tension) are characterized by the greatest impact of the Pd content. Such impact may dominate the transformation behaviour and obscure entirely the role of the Pd content in terms of phase compatibility and transformation volume change. We thereby limit our main analysis to the [111] orientation, which exhibits the lowest impact of the Pd content on the transformation Schmid factors.

Fig.~\ref{Fig-NiTiPdMicFD} shows the effect of the Pd content on the $P$--$\delta$ response and microstructure at the maximum indentation depth $\delta_\text{max}=28$ nm. We stress again that the stretch parameters ($\alpha$, $\beta$, $\gamma$), and thus the transformation stretch tensors $\bfm{U}_i^\text{t}$, are the only varying input parameters,  while the dependence of elastic constants, interfacial energy, and other parameters on the Pd content is neglected, as the related data is not available. At the first glance, the simulation results suggest a negligible influence of the Pd content. Despite some small differences between individual microstructures, see Fig.~\ref{Fig-NiTiPdMicFD}(c), a consistent and meaningful trend of microstructure change is hardly visible. A quantitative examination based on the $P$--$\delta$ responses, however, reveals that the maximum indentation load (associated with $\delta_\text{max}$) and the enclosed area of the pseudoelastic loop are directly influenced by the Pd content, although the related differences are not large (less than 10\%). Fig.~\ref{Fig-NiTiPdAnal}(b) presents the detailed results in terms of the hysteresis and hysteresis/work ratio as a function of the Pd content. Herein, hysteresis is defined as the enclosed area of the pseudoelastic loop (equal to the energy dissipated in the complete forward--reverse transformation cycle), and the work is defined as the area below the loading branch. Interestingly, the graphs in Fig.~\ref{Fig-NiTiPdAnal}(b) demonstrate the maximum hysteresis for 11\% Pd, which is characterized by $\lambda_2=1$ and by the largest transformation volume change. To establish a meaningful correlation, the hysteresis data are plotted in Fig.~\ref{Fig-NiTiPdAnal}(c) as a function of the transformation volume change $J_\text{t}$. It can be seen (in particular for the hysteresis/work ratio) that the graphs have a non-monotonic trend with the minimum at $J_\text{t}=0.9942$ (for 20\% Pd), showing the optimum contraction that leads to the lowest transformation hysteresis during nanoindentation. Note that indentation induces compressive stresses beneath the indenter, hence a negative volume change ($J_\text{t}<1$), as in NiTiPd, should be favoured by the stress state.

Further analysis of the results reveals that no relevant connection can be established between the hysteresis data and the second eigenvalue $\lambda_2$. This can be justified as follows. According to the crystallographic theory, the effect of $\lambda_2$ is manifested through the change in the microstructure pattern of planar austenite--martensite interfaces. Specifically, twinned microstructure is developed for $\lambda_2 \neq 1$ and twinless microstructure for $\lambda_2=1$. In the present simulations, the size of the transformed domain is rather small. Therefore, the interfacial energy acts as the main factor that governs the microstructure evolution, and, in particular, prevents development of planar austenite--martensite interfaces. As a result, the $\lambda_2$ indicator does not show the expected impact on the microstructure and on the hysteresis.

\section{Conclusions}\label{sec-concl}
This paper presents the first attempt to use the phase-field method for a physically-insightful 3D analysis of martensitic transformation during nanoindentation. A finite-element-based phase-field model, developed recently by the authors \citep{tuuma2021phase}, is employed for this purpose. The focus of the study is on pseudoelastic SMA single crystals, namely CuAlNi and NiTiPd.

A pronounced orientation dependence of the mechanical response and microstructure has been captured for CuAlNi. A comprehensive investigation of the microstructures revealed that: (i) consistent with the experiments, the evolved twin laminates are either type II or compound twins, (ii) the orientation of the twin planes is in agreement with those calculated by the crystallographic theory of martensite, and (iii) the selection of the martensite variants can be reasonably explained by a simple Schmid factor analysis, all demonstrating the credibility of the simulation results. 

No appreciable differences have been detected in the microstructures predicted for NiTiPd with different Pd contents, despite the visible differences in the transformation stretch parameters. It has been found out that the maximum and minimum transformation hysteresis in nanoindentation are obtained for the Pd content of 11\% and 20\%, respectively. This peculiar observation implies the predominant influence of the transformation volume change $J_\text{t}$ over that of the middle eigenvalue $\lambda_2$ and is in contrast with the general consensus reached by the experimental observations (in temperature-induced transformation) and advocated by the geometrically nonlinear crystallographic theory of martensite. It can be argued that the overwhelming impact of the interfacial energy, which stems from the limited size of the computational domain and, thereby, the relatively small size of the transformed domain, explains the failure of the middle eigenvalue $\lambda_2$ in inducing a meaningful impact on the microstructures and on the transformation hysteresis.

Our study highlights also the key role of some specific model features, including elastic anisotropy and finite-strain kinematics. It follows that both features significantly impact the final results. More importantly, the finite-strain kinematics proves to be a crucial element for a reliable prediction of the microstructure pattern.

\paragraph{Acknowledgement} MRH has been supported by the National Science Centre (NCN) in Poland through Grant No.\ 2021/43/D/ST8/02555. KT has been supported by the ERC-CZ grant LL2105 of the Ministry of Education, Youth and Sport of the Czech Republic and by Charles University Research program No.\ UNCE/SCI/023. SS has been supported by the National Science Centre (NCN) in Poland through Grant No.\ 2018/29/B/ST8/00729. This work has been also supported by the Ministry of Education, Youth and Sports of the Czech Republic through the e-INFRA CZ project (ID:90140).

\bibliographystyle{unsrtnat}
\biboptions{sort&compress}

\bibliography{bibliography}

\end{document}